\documentclass[journal=jacsat,manuscript=article]{achemso}

\usepackage{amssymb}
\usepackage{amsmath}
\usepackage[labelsep=period]{caption}

\usepackage{hyperref}
\hypersetup{
    colorlinks,%
    citecolor=blue,%
    linkcolor=blue,%
    urlcolor=blue
}
\usepackage{chemformula} 
\usepackage[T1]{fontenc} 
\author{Dai Q. Ho}
\affiliation{Department of Materials Science and Engineering, University of Delaware, Newark, DE 19716, USA}
\alsoaffiliation{Faculty of Natural Sciences, Quy Nhon University, Quy Nhon 590000, Vietnam}
\email{daiqho@udel.edu}
\author{Ruiqi Hu}
\affiliation{Department of Materials Science and Engineering, University of Delaware, Newark, DE 19716, USA}
\author{D. Quang To}
\affiliation{Department of Materials Science and Engineering, University of Delaware, Newark, DE 19716, USA}
\author{Garnett W. Bryant}
\affiliation{Nanoscale Device Characterization Division, Joint Quantum Institute, National Institute of Standards and Technology, Gaithersburg, Maryland 20899-8423, USA and University of Maryland, College Park, Maryland 20742, USA}
\email{garnett.bryant@nist.gov }
\author{Anderson Janotti}
\affiliation{Department of Materials Science and Engineering, University of Delaware, Newark, DE 19716, USA}
\email{janotti@udel.edu}
\title[An \textsf{achemso}]
  {Non-trivial topology in rare-earth monopnictides from dimensionality reduction}


\begin{document}


\begin{abstract}
    Thin films of rare-earth monopnictide semimetals are expected to turn into semiconductors due to quantum confinement effect, which lifts the overlap between electron pockets at Brillouin zone edges and hole pockets at the zone center. Instead, taking non-magnetic LaSb as an example, we find the emergence of a quantum spin Hall insulator phase in LaSb(001) films as the thickness is reduced to 7, 5, or 3 monolayers. This is attributed to a strong quantum confinement effect on the in-plane electron pockets, and the lack of quantum confinement on the out-of-plane pocket in reciprocal space projected onto zone center, leading to a band inversion. Spin-orbit coupling opens a sizeable non-trivial gap in the band structure of the thin films. Such effect is shown to be general in rare-earth monopnictides and may lead to interesting phenomena when coupled with the 4$f$ magnetic moments present in other members of this family of materials.
\end{abstract}
\begin{figure}
\includegraphics[width=1\textwidth]{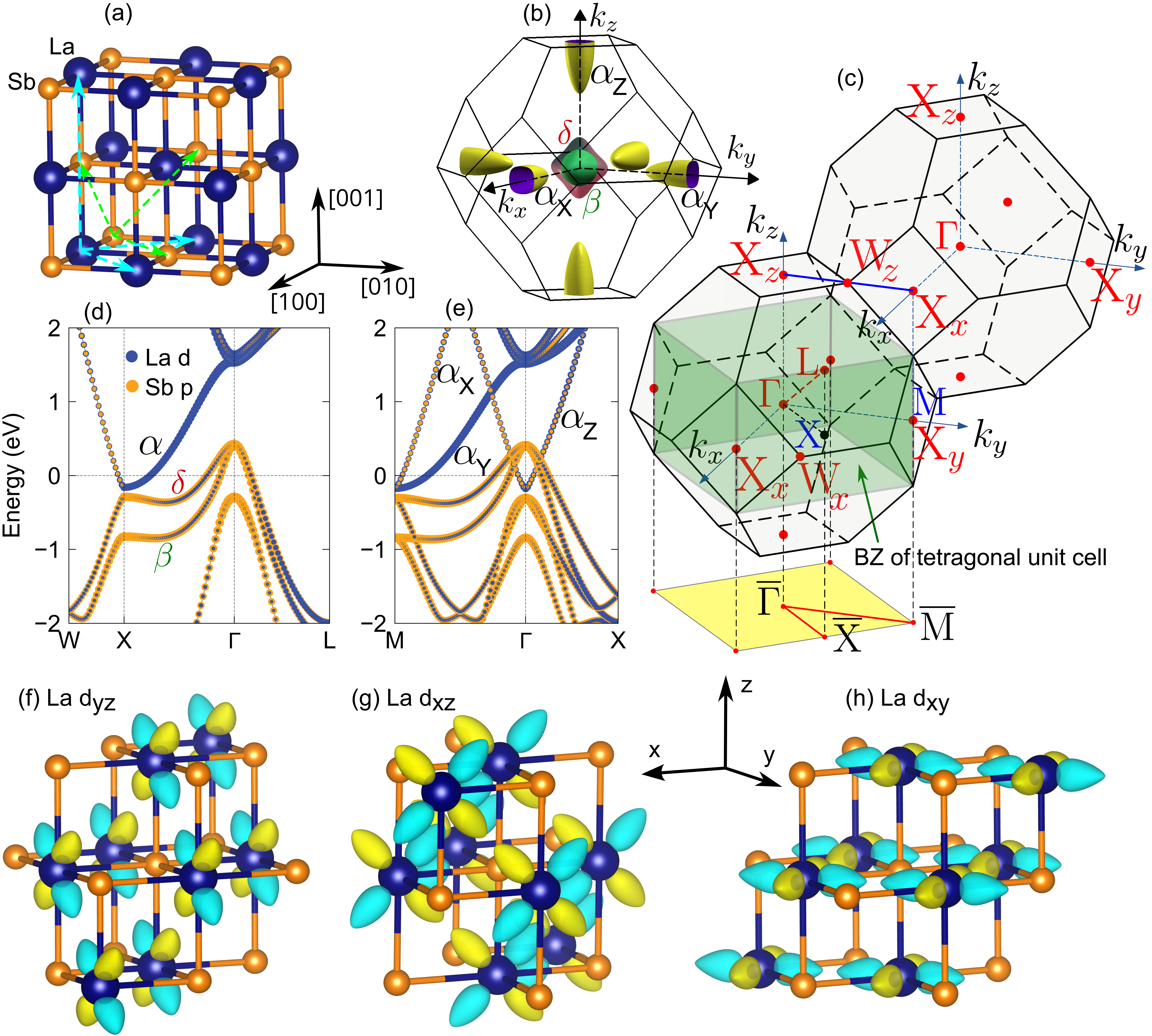}
\caption{\label{fig1}Electronic structure of a typical semimetallic RE-V. (a) Conventional rock-salt crystal structure, the green and cyan arrows representing the unit vectors of bulk primitive unit cell and tetragonal unit cell, respectively, (b) Fermi surface of the primitive bulk semimetallic RE-V featuring two hole pockets at ${\Gamma}$ and three equivalent electron pockets at X points, (c) 3D BZ of RE-V $fcc$ crystal structure and its projection on (001) plane (yellow plane) and also shown is the 3D BZ of the tetragonal unit cell (right square green prism inside the BZ of the primitive $fcc$); orbital-resolved band structure of semimetallic RE-V using (d) primitive unit cell, (e) 4-atom tetragonal unit cell at $k_z=0$, more bands observed than that of the primitive unit cell due to band folding phenomenon; visualization of the wavefunctions of the electron pockets centered at (f) X$_x$, (g) X$_y$, and (h) X$_z$ (yellow and cyan lobes represent $+$ and $-$ sign of the wavefunctions, respectively).}
\end{figure}

Rare-earth monopnictides (RE-Vs), with simple rock-salt crystal structure composed of two interpenetrating face-centered cubic ($fcc$) sublattices (Figure~\ref{fig1}a), are fully compensated semimetals with electron pockets centered at three X points and two hole pockets at $\Gamma$, as shown in Figure~\ref{fig1}b for LaSb. RE-V thin films have been epitaxially grown on III-V semiconductors \cite{palmstrom1988epitaxial, buehl2010growth, kawasaki2013surface} and sought as structurally perfect (defect-free) epitaxial contacts \cite{palmstrom1988epitaxial,hanson2006eras}, as nanoparticles embeded in III-Vs for thermoelectrics, infrared and terahertz generation and detection \cite{liu2011properties,kawasaki2011local,kawasaki2013size, lu2014self,bomberger2017overview,krivoy2018rare}. Some members of this material family display interesting properties such as extremely large magneto-resistance and non-trivial topological phase including a recently observed unusual magnetic splitting phenomenon \cite{tafti2016resistivity,zeng2016LaSbcompensated,wu2016asymmetric,nayak2017multiple,lou2017evidenceLaBi,kuroda2018experimentalCePn,schrunk2022emergence}. The relatively small overlap in energy between the electron and hole pockets have led researchers to expect that this overlap could be lifted and a band gap would ultimately be opened in structures with reduced dimension such as nanoparticles or thin films due to the quantum confinement effect (QCE). So far, this has not been realized, and the reasons for it have remained unknown \cite{allen1990magneto,kawasaki2011local,kawasaki2013size}.

Using first-principles calculations for free standing slabs of LaSb (taken as an example, here) we show that the overlap in energy between the electron pockets lying in the film-plane directions (X$_x$ and X$_y$) and the hole pockets at $\Gamma$ indeed decreases upon dimensional reduction in thin films, but the electron pocket along the film-plane normal direction, at X$_z$, which is projected onto ${\overline{\Gamma}}$ in the 2D Brillouin zone (BZ) of the thin film, remains. This direction dependent behavior is attributed to the directional nature of the electron pocket wavefunctions, originating from the rare-earth 5$d$ orbitals, which maintains the metallic character in thin films. However, for films that are 7 or less monolayers (ML) thick, we show that orbital-selective QCE may ultimately lead to an inverted band gap which produces helical spin-momentum locked edge states in quasi 2D nanoribbons. Calculations of the $\mathbb{Z}_2$ invariant, spin-resolved local density of states (LDOS), and spatial distribution of wavefunctions associated with edge modes confirm the non-trivial topology in the electronic structure of these ultra-thin films. This material system provides a novel platform for realizing the intrinsic quantum spin Hall (QSH) insulator phase in experiment different from the two well-known examples of semiconductor quantum wells \cite{bernevig2006quantum,konig2007quantum} and monolayer transition metal dichalcogenides \cite{qian2014quantum, tang2017quantum, fei2017edge, wu2018observation}, as well as some other recent proposals \cite{tang2014stable,liu2017van,kandrai2020signature}.

The calculated lattice parameter of LaSb is 6.547 \AA~and 6.518 \AA~using the DFT-GGA and the HSE06 hybrid functional, respectively, both in good agreement with previous calculations \cite{Shoaib2020} and the experimental value of 6.488 \AA~\cite{chua1974simple} which is adopted in this study. The computed band structure of primitive bulk LaSb is shown in Figure~\ref{fig1}d. There are three electron pockets (denoted by $\alpha$ centered at X$_x$, X$_y$, and X$_z$) from bands derived mainly from La 5$d$ orbitals, and two hole pockets ($\beta$ and $\delta$ centered at $\Gamma$) from bands derived from Sb 4$p$ orbitals, crossing the Fermi level, leading to full compensation, i.e., the total numbers of electrons and holes are equal. This overlap is overestimated in DFT-GGA, resulting in an artificially higher carrier concentration compared to experimental values, while HSE06 gives carrier concentration in better agreement with experiments \cite{Shoaib2020}. As can be seen in Figure~\ref{fig1}b, the electron pockets are highly anisotropic with each ellipsoid's semi-major axis pointing along the corresponding Cartesian axis, and the semi-minor axis being perpendicular to it. The hole pockets, however, are more isotropic with an inner spherical-like pocket $\beta$ and an outer warped double pyramid pocket $\delta$. Therefore, we expect that the electron pockets will be more sensitive to QCE upon dimensional reduction in thin films than the hole counterparts.

The effect of quantum confinement on the electron pockets can be understood by inspecting the single-particle wavefunctions at the three X points shown in Figure~\ref{fig1}f-h. At X$_x$ (X$_y$), the wavefunction is composed of La $d_{yz}$ ($d_{xz}$) orbital (see details on projected band structure in Figure S1), resulting in a low dispersion along the semi-major axis $\Gamma-$X$_x$ ($\Gamma-$X$_y$) in reciprocal space because the orbital interaction occurs along the direction normal to their charge distribution as evidenced in Figure~\ref{fig1}f(g). Meanwhile, the same band has a higher dispersion along the perpendicular direction, i.e., the electron pocket's semi-minor axis X$-$W thanks to the orbital interaction happening in the plane of their distribution ($yz$ and $xz$ planes for pockets at X$_x$ and X$_y$, respectively).
At X$_z$, on the other hand, the wavefunction is composed of La $d_{xy}$ orbital, i.e., lying in the $xy$ plane with small overlap along the [001] or $z$ direction (Figure~\ref{fig1}h). Therefore, we anticipate that QCE in [001] oriented thin films will affect mostly the electron pockets whose wavefunctions are largely distributed in the planes normal to the film plane, i.e., the pockets centered at X$_x$  and X$_y$; the electron pocket at X$_z$ will be marginally affected by the QCE because the corresponding wavefunction is spreading in the plane of the film \cite{chang2009fundamental,sohn2021observation}. This can be checked by examining the band structure of thin free-standing slabs.

To understand the electronic structure of [001] oriented films, we first discuss the band structure of LaSb in the tetragonal unit cell, with two formula units, at $k_z=0$, as shown in Figure~\ref{fig1}e which has similar symmetry to that of the (001) films, i.e., we can build a (001) film unit cell from the tetragonal unit cell. Its corresponding BZ is highlighted by the green right square prism inside the BZ of the $fcc$ primitive cell in Figure~\ref{fig1}c.
The electron pocket at $\Gamma$, overlapping in momentum and energy with the hole pockets, corresponds to the pocket at X$_z$ due to the folding of the BZ along $k_z$ (see details from projected band structure in Figure S2). It can also be viewed as the projection on the $k_x$-$k_y$ plane of the ellipsoid along $k_z$ in the Fermi surface of the primitive cell (Figure~\ref{fig1}b).  Note that the overlap and crossing of electron and hole pockets near $\Gamma$ in Figure~\ref{fig1}e is a consequence of having a larger unit cell compared to the primitive cell. Otherwise, they would open an anti-crossing gap in the presence of spin orbit coupling (SOC). In fact, the electron pocket at X$_z$ and hole pockets at $\Gamma$ belong to distinct symmetry irreducible representations (irreps), and therefore, the interaction between them is prohibited by symmetry, i.e., the C$_4$ rotation operation along $\Gamma-$X$_z$\cite{kim2018nearly,li2015gdn}. However, for finite thickness films along [001], this symmetry is lifted, and the overlap in both energy and momentum becomes viable \cite{li2015gdn}.
At the M point, there are two electron bands with different dispersion: the lower dispersion band composed of La $d_{xz}$, originally from the pocket at X$_y$, with the dispersion along the semi-major axis $\Gamma-$X$_y$, and the higher dispersion counterpart originating from the projection of the semi-minor axis X$_x-$W$_z$ of the pocket at X$_x$ from the adjacent BZ (the path highlighted in blue in Figure~\ref{fig1}c).
To summarize, the three electron pockets of the bulk in primitive unit cell form are residing in two locations of the tetragonal unit cell BZ: X$_z$ at $\Gamma$, X$_x$ and X$_y$ at M, which would ultimately determine the effect of size quantization upon dimensional reduction.

\begin{figure}
\centering
\includegraphics[width=1.0\textwidth]{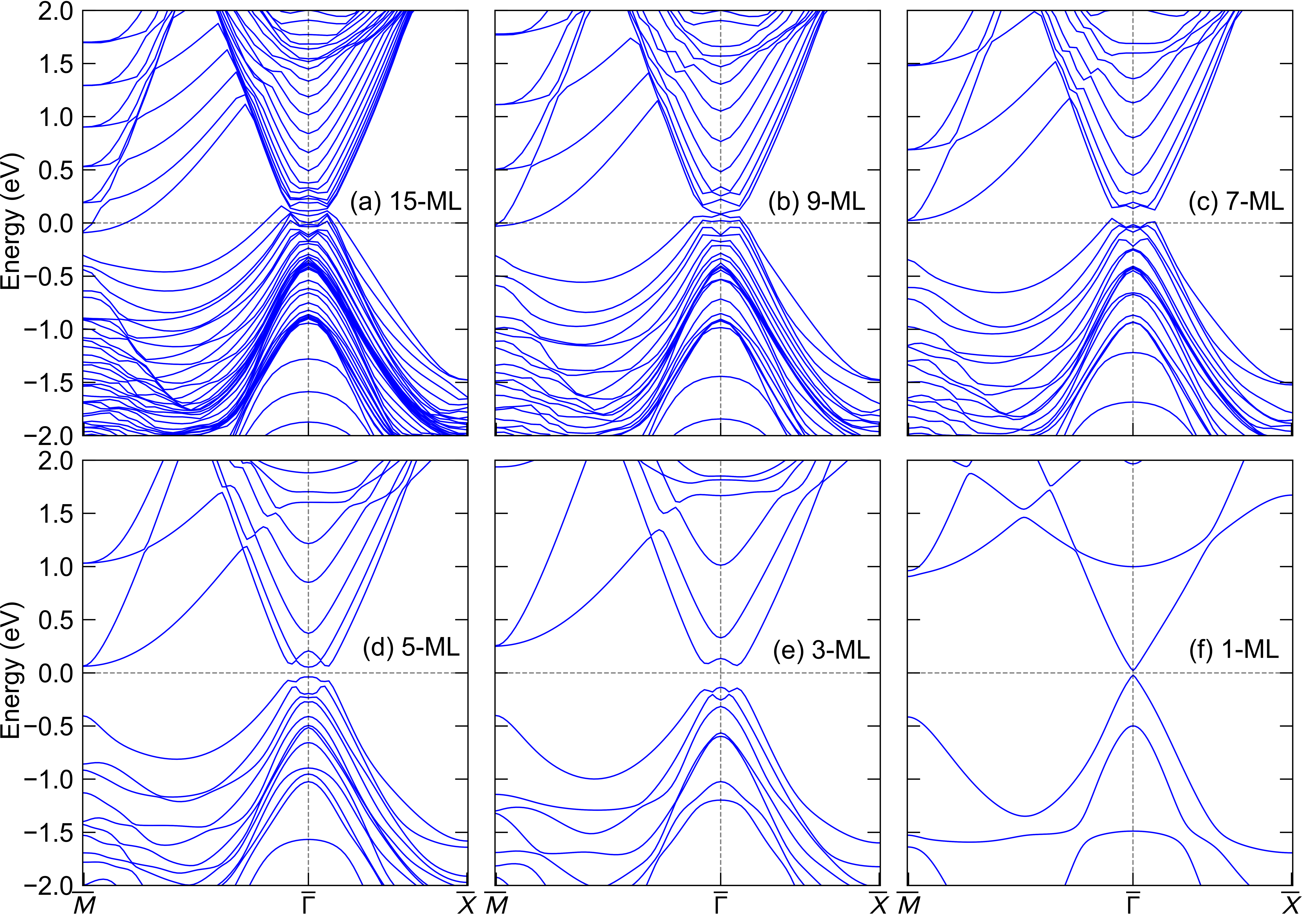}
\caption{\label{fig2}Electronic structure of freestanding films LaSb with the thickness of (a) 15-ML, (b) 9-ML, (c) 7-ML, (d) 5-ML, (e) 3-ML, (f) 1-ML. Due to the orbital-selective QCE on the electron pockets at $\overline{\rm M}$ and $\overline{\Gamma}$, a phase transition from topologically trivial semimetal in 15-ML film to non-trivial insulator in 7-, 5-, 3-ML films and eventually trivial insulator in 1-ML film can be observed.}
\end{figure}

The electronic band structure of a free standing slab with 15 MLs is shown in Figure~\ref{fig2}a, along $\overline{\rm M}-\overline{\Gamma}-\overline{\rm X}$, where $\overline{\rm M}$ can be viewed as the projection of the two high symmetry points X$_y$ and X$_x$ (from the adjacent BZ), $\overline{\Gamma}$ is the projection of $\Gamma$ and X$_z$, and $\overline{\rm X}$ is the projection of the L points onto the 2D BZ of the thin films (Figure~\ref{fig1}c). We note that the electron pocket at $\overline{\rm M}$ (corresponding to X$_x$ and X$_y$) is shifted upwards in energy while the electron pocket at $\overline{\Gamma}$ (projected from X$_z$) remains nearly unchanged. At the $\overline{\rm M}$ point, there are two sets of electron pocket sub-bands with quite different dispersion: one set having low dispersion (high effective mass) while the other set exhibiting large dispersion (low effective mass). This can be understood from the orbital character of the bands (Figure S3) and the directional nature of their interaction as analyzed above for the M point of the tetragonal unit cell case. The low dispersion set of sub-bands stems from the La $d_{xz}$ orbital interaction taking place along the semi-major axis $\Gamma-$X$_y$. Meanwhile, the high dispersion set of sub-bands originates from the semi-minor axis X$_x-$W$_z$ due to the in-plane interaction of La $d_{yz}$, resulting in a stronger interaction and hence a smaller effective mass. The gaps at the X$_x$ and X$_y$ points in the bulk transfer to the gap at the $\overline{\rm M}$ point of the film. At $\overline{\Gamma}$, however, electron pockets overlaps with hole pockets, maintaining the semimetallic phase at this thickness of the film. In addition, the energy separation of the sub-bands at $\overline{\rm M}$ is associated with the dispersion, or effective mass, of the electron band at X$_x$ (X$_y$). Consequently, this energy separation is larger than that between the sub-bands of the electron pockets at $\overline{\Gamma}$ due to QCE; the latter is associated with the coupling of the $d_{xy}$ orbitals in the film plane, hence less affected by QCE.

The electronic structure of the 9-ML thick slab is shown in Figure~\ref{fig2}b, indicating a semimetal on the brink of having the electron pocket at $\overline{\rm M}$ above the Fermi level. For the 7-ML and 5-ML thick slabs, Figures~\ref{fig2}c,d show that the electron pockets at $\overline{\rm M}$ are above the Fermi level due to the QCE, and inverted band gaps at $\overline{\Gamma}$ are opened in the presence of SOC, indicating the emergence of a 2D (or quasi 2D) QSH phase \cite{kane2005quantum,sheng2005nondissipative,bernevig2006quantum}. The gaps at $\overline{\rm M}$ and at $\overline{\Gamma}$ are enlarged in the 3-ML thick film (Figure~\ref{fig2}e). At the ultimate limit of 1-ML thick, LaSb film becomes a normal insulator thanks to the QCE lifting up the overlap (at $\overline{\Gamma}$) between the projected electron pocket from $X_z$ and hole pockets, leading to a normal band ordering (Figure~\ref{fig2}f).  

\begin{figure*}
\centering
\includegraphics[width=1.0\textwidth]{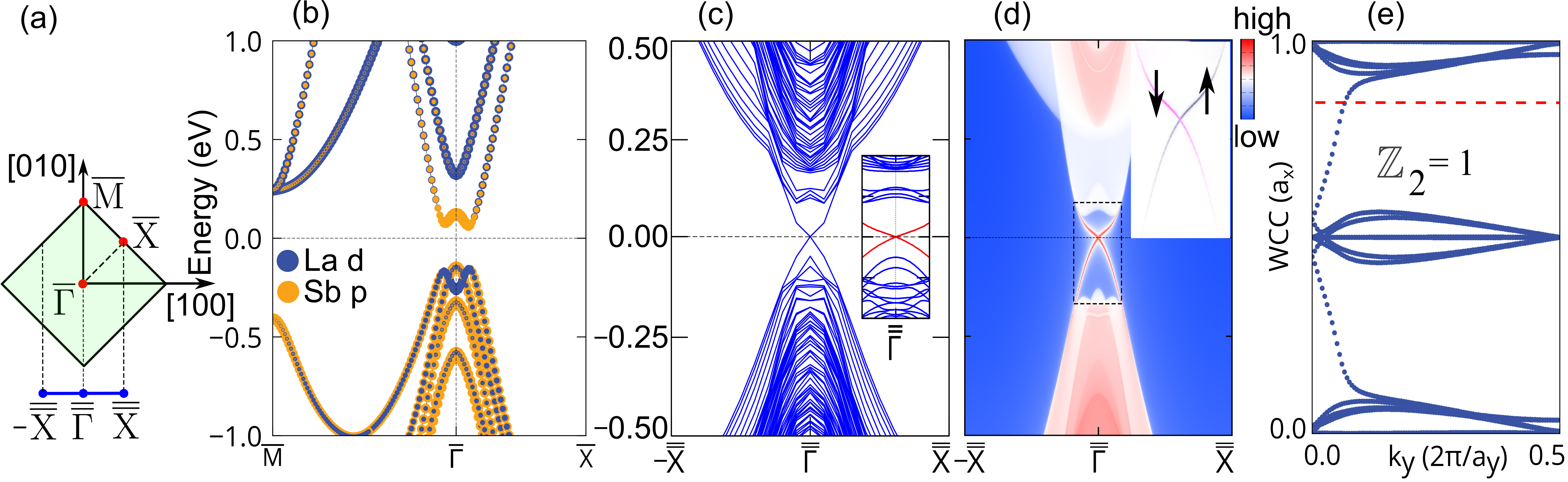}
\caption{\label{fig3}Signature of the QSH insulating phase in 3-ML thick film. (a) 2D BZ of the thin film and 1D BZ of the ribbon made of the film (blue line) with periodic direction along $x$, (b) projected band structure around the Fermi level (set at 0 eV) exhibiting an inverted band gap at $\overline{\Gamma}$, (c) electronic structure of the ribbon made from the 3-ML thick film obtained from $ab$ $initio$ calculations using openMX showing the presence of edge states crossing at Fermi level as highlighted by red lines in the inset, (d) LDOS of a semi-infinite wide ribbon made from the 3-ML thick LaSb obtained from Wannier orbital-based TB Hamiltonian and the spin-polarized LDOS of the edge states shown in the inset, (e) Wannier charge center evolution of the 3-ML thick LaSb freestanding film displaying the non-trivial topological property as the red dashed reference line crossing the evolution line (represented by filled navy blue circles) an odd number of time.}
\end{figure*} 

As mentioned above, the electronic structure of LaSb films evolves from a semimetal with trivial topology to a non-trivial insulator as the film thickness decreases down to a few MLs. The topologically non-trivial electronic structure follows a band inversion, as shown in Figure~\ref{fig3}b for the case of 3-ML thick film. To confirm the topological non-triviality, we calculated the $\mathbb{Z}_2$ topological invariant using the Fu-Kane formula \citep{fu2007topological} as well as the Fukui-Hatsugai method \citep{fukui2005chern}, obtaining $\mathbb{Z}_2=1$. The non-trivial topological nature is further corroborated by following the evolution of the Wannier charge center (WCC) as a function of momentum, shown in Figure~\ref{fig3}e, which essentially displays the non-trivial topology as the red-dashed reference line cutting through the WCC evolution line an odd number of times. 

Furthermore, based on the bulk-boundary correspondence, we expect edge states of the QSH insulator to show the spin-momentum locking behavior \cite{kane2005quantum}. We  searched for the presence of spin-polarized edge modes associated with the non-trivial topology by calculating the band structure of a nanoribbon made from the 3-ML film. We find that the nanoribbon must be at least 50 MLs wide ($\sim$16 nm) to avoid interactions between the two opposite edges of the ribbon. The electronic structure of the 3-ML thick, 50-ML wide nanoribbon is shown in Figure~\ref{fig3}c, where the edge states cross at the four-fold degenerate Dirac point $\overline{\overline{\Gamma}}$, represented by a pair of red lines in the vicinity of $\overline{\overline{\Gamma}}$ in the inset. The corresponding spin-resolved LDOS in Figure~\ref{fig3}d, obtained by the iterative Green function method \cite{sancho1985highly} for a semi-infinite structure, and the spatial distribution of spin-resolved wavefunctions of edge states shown in Figure S4, clearly demonstrate the helical spin-momentum locking of the edge modes, thus confirming the presence of the QSH phase in the 3ML-thick film.

For 7-ML or 5-ML thick films, there are two set of band inversions (see Figure S5a,e), thus no time-reversal symmetry-protected edge states are expected, i.e., spin scattering from spin up to spin down channels on the same edge is possible \cite{qi2010quantum,maciejko2011quantum}. Therefore, they are classified as trivial insulators with the topological invariant $\mathbb{Z}_2=0$ \cite{}. However, one pair of the inverted bands can be removed when a small biaxial tensile strain is applied to films deposited on an appropriately chosen substrate. Such an approach has been investigated theoretically \cite{khalid2020trivial} and recently demonstrated experimentally in the case of RE-Vs grown on III-V substrates \cite{inbar2022strain}. This biaxial (or epitaxial) strain allows to selectively lift the overlap between one pair of electron and hole sub-bands, resulting in an odd number of band inversions, so that a QSH phase can also be realized in 5-ML and 7-ML thick films (see Figure S5c,d,h). 

\begin{figure}
\centering
\includegraphics[width=0.5\textwidth]{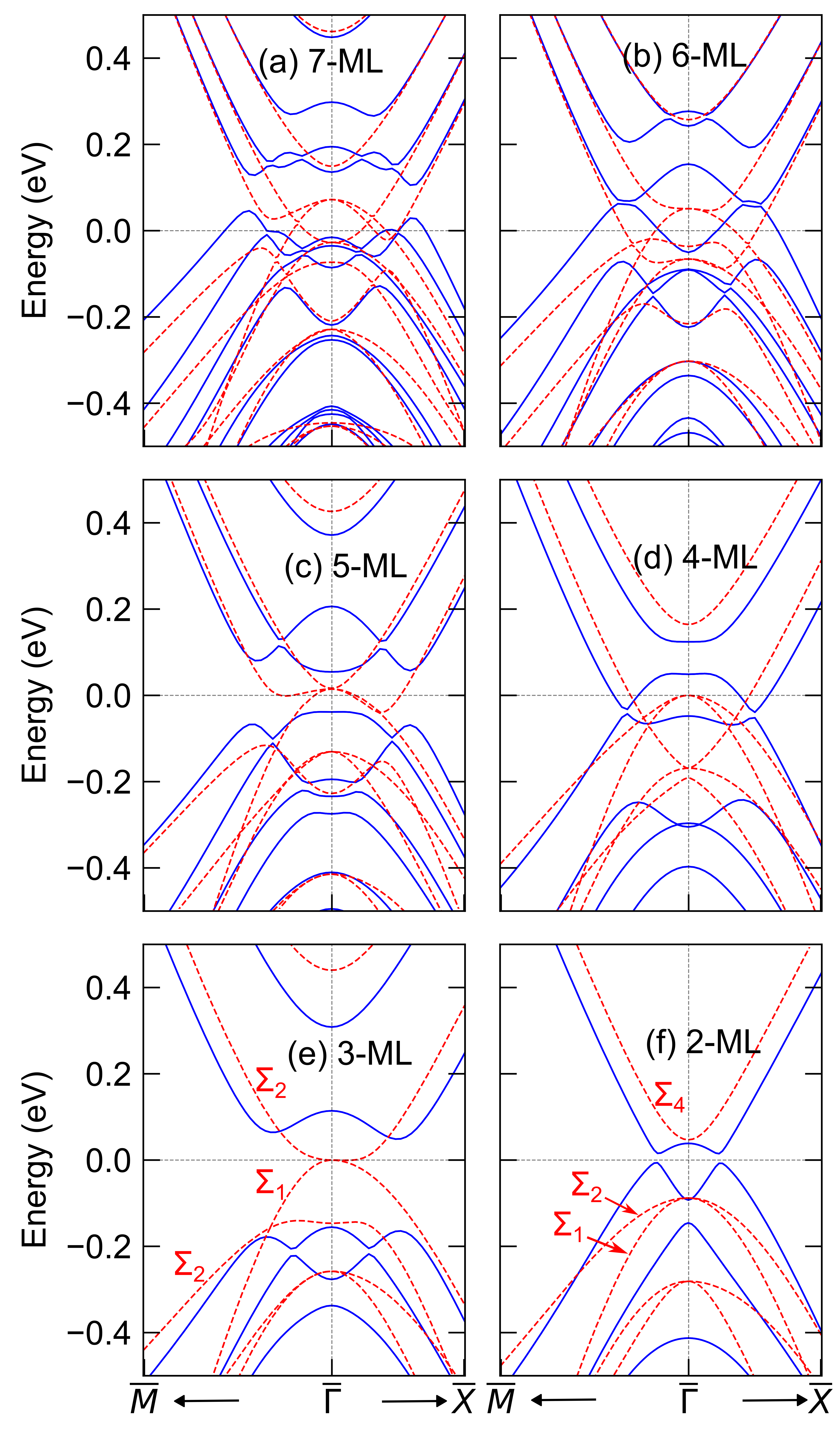}
\caption{\label{fig4}Electronic structures in the proximity of the $\overline{\Gamma}$ point of LaSb freestanding thin films with an even/odd number of MLs. Red dashed and blue lines indicate bands without and with SOC, respectively. Irreps of relevant bands are denoted by capitalized Greek letters.}
\end{figure}

When taking a closer look at the thickness-dependence of the electronic structure of the thin films, there is a striking difference between films having even and odd number of monolayers (Figure~\ref{fig4} and Figure S6), particularly near the Fermi level. As the thickness of the film decreases, the QCE becomes stronger, resulting in gradually larger sub-band. The QCE also affects the size of the gap at the $\overline{\rm M}$ and $\overline{\Gamma}$ points. Thinner films with an odd number of MLs show larger gaps with decreasing film thickness. The same trend is seen for films with an even number of MLs. Remarkably, the gap at $\overline{\Gamma}$ oscillates when going from an even to an odd number of ML, as detailed in Figure~\ref{fig4} for the bands around $\overline{\Gamma}$. The gap is essentially zero for films having an even number of MLs but becomes sizable for films with an odd number of MLs.

This oscillation in the gap size of the thin films with even and odd number of MLs can be explained by the difference in their symmetry. The films having an odd number of MLs belong to a symmorphic symmetry group (space group number 123 with point group $P4/mmm$) while the films with an even number of MLs possess a nonsymmorphic symmetry group (space group number 129 with point group $P4/nmm$). This leads to a distinction in the symmetry of the bands around the Fermi levels as shown in Figure~\ref{fig4}e,f taking 3-ML and 2-ML films as examples. For the 3-ML film, one electron band and two hole bands overlap in the proximity of $\overline{\Gamma}$, leading to a band inversion at $\overline{\Gamma}$ in the absence of SOC (represented by red dashed lines). Crucially, one of the hole bands has the same symmetry as that of the electron band (irrep $\Sigma_2$, Figure~\ref{fig4}e), leading to a gap opening due to their mutual interaction. However, the other hole band with a different irrep ($\Sigma_1$) does not interact with the electron counterpart, resulting in a semimetallic phase with a degeneracy of the two hole bands at the $\overline{\Gamma}$ point. This inverted band gap semimetal with a degenerate point at the zone center is similar to bulk HgTe \cite{bernevig2006quantum}. The degeneracy in bulk HgTe is protected by its crystal symmetry and is stable even in the presence of SOC; it is only lifted and a tiny gap opens when the bulk crystal symmetry is broken, i.e., in the thin film structure (quantum well). For the 3-ML LaSb film, however, the degeneracy is not stable under the influence of SOC since it is from the original degeneracy of the top two hole bands at the X$_z$ point of the bulk when not taking SOC into consideration. As a consequence, when turning on SOC, a sizeable inverted gap comparable to the SOC splitting between $\beta$ and $\delta$ bands at the bulk X point is observed.

On the other hand, for the 2-ML system the two hole bands and the electron band have different symmetries, characterized by irreps $\Sigma_1$, $\Sigma_2$, and $\Sigma_4$, respectively (Figure~\ref{fig4}f). Therefore, in the absence of SOC, those bands would cross if they were overlapping in energy since their interactions are prohibited by symmetry (as in the case of 4ML-thick film shown in Figure~\ref{fig4}d). When SOC is turned on, the highest hole band and the electron band overlap in energy, leading to a SOC-induced gap opening. However, this gap is tiny since the SOC plays a role as a small perturbation, the relevant orbitals interaction is not allowed by symmetry. A similar argument can be applied to understand the sizeable inverted gap in the other films with odd numbers of MLs such as the 5-ML and 7-ML, as well as the observed essentially zero gaps in the films with an even number of MLs such as the 4-ML and 6-ML films.

The effects discussed here are expected to occur for all materials in the family of RE-Vs with RE=La,..., Lu and V=As, Sb, Bi since they have the same crystal structure and similar orbital composition of the bands.  Note, however, that some of the Bi-based compounds already show non-trivial topology in the 3D bulk due to the crossing of the Bi 6$p$ and the rare-earth 5$d$ bands near the X point \cite{lou2017evidenceLaBi,nayak2017multiple}.  The As- and Sb-based compounds are all topologically trivial semimetals, and hence are all expected to become topologically non-trivial when made as few monolayers thick, [001] oriented ultra-thin films. Apart from the case of La-, Y-, and Lu-based compounds, where the RE $f$ orbitals are completely empty or filled, other RE-Vs exhibit magnetic ordering at low temperatures, opening an avenue to combining non-trivial topology with spin magnetism, potentially leading to novel phenomena such as the emergence of 2D antiferromagnetic topological insulators \cite{niu2020antiferromagnetic, liang2022approaching, xu2022evolution}, the quantum anomalous Hall effect in Chern insulators \cite{inoue2019band}, and Fermi arcs due to an unusual magnetic splitting \cite{schrunk2022emergence}. 

Experiments have already demonstrated the epitaxial growth of RE-Vs thin films on conventional III-V substrates, down to few MLs thick. Scanning tunneling spectroscopy measurements on GdSb/GaSb has explored the possibility of turning GdSb semimetal into a semiconductor in ultra-thin films, yet failing \cite{inbar2022epitaxial,inbar2022strain}. The authors, however, were not searching for edge states at that time. We also argue that the presence of metallic interface states at the interface with III-Vs \cite{chatterjee2021controlling} might impose difficulties in identifying the spin-momentum locked states and that perhaps a different substrate with different bonding at the interface might be required. In the case of LaSb, we propose the use of rock-salt MgTe, CaTe, or SrTe for which interface metallic states would be mitigated, facilitating the observation of the spin-momentum locked edge states.

In conclusion, we have investigated the electronic structure of non-magnetic RE-V thin films at various thicknesses using first-principles calculations employing the hybrid functional HSE06, taking LaSb as a case study. The quantum confinement in thin films along [001] has dissimilar effects on the electron pockets centered at equivalent X points at the zone edges owing to their orbital composition. The directional nature of the wavefunction associated with the electron pocket at X$_z$ point, distributed mainly in the film plane with weak out-of-plane interactions, makes this pocket resistant to finite size effects, leading to an inverted band gap opening in the limit of very thin films. Interestingly, at the limit of 3-ML, we have observed the emergence of a QSH insulating phase characterized by the non-trivial $\mathbb{Z}_2$ topological invariant and the presence of helical spin-momentum locked edge states. Similar observation can be seen for 5-ML and 7-ML films with an applied small biaxial tensile strain. Our results indicate that this effect is general for the RE-Vs, with a possibly interesting combination of 4$f$ magnetism with topological band structures in ultrathin films.

\section{Computational methods}

First-principles calculations based on density functional theory \cite{Hohenberg1964, KohnSham1965} as implemented in the VASP code \cite{kresse1996efficient} have been performed for structural optimization and electronic structure calculation. The screened hybrid exchange-correlation functional of Heyd-Scuseria-Ernzerhof (HSE06) \cite{hse06-2003,hse06} was employed in all cases except the calculations of the electronic structures of nanoribbons because these calculations required very large unit cells. In that case, the generalized gradient approximation (GGA) functional of Perdew–Burke-Ernzerhof (PBE) \cite{Perdew1996-PBE,Perdew1997-PBE-Erratum} was used instead. A plane-wave basis set with up to kinetic cutoff energy of 400 eV, and 12$\times$12$\times$12 $k$-point mesh for bulk LaSb and 10$\times$10$\times$1 $k$-point mesh for thin films were used, respectively. Due to large unit cell sizes of the LaSb nanoribbons, first-principles calculations searching for the presence of edge states were performed using the OpenMX code \cite{Ozaki2003,Ozaki2004,Ozaki2005} with a localized basis set. The symmetry irreducible representations (irreps) of electronic bands were determined using the irvsp code \cite{gao2021irvsp}. Maximally localized Wannier functions (MLWFs) based effective Hamiltonian were constructed by the Wannier90 code \cite{pizzi2020wannier90} using La $d$ and Sb $p$ as projectors. Edge state spectra were then obtained by using the surface Green function for semi-infinite systems \cite{sancho1985highly} as implemented in WannierTools \cite{wu2018wanniertools}.

\begin{acknowledgement}
This research was supported by the NSF through the UD-CHARM University of Delaware Materials Research Science and Engineering Center (No. DMR-2011824).  It made use of the computing resources provided by the Extreme Science and Engineering Discovery Environment (XSEDE), supported by the National Science Foundation grant number ACI-1053575.
\end{acknowledgement}

\begin{suppinfo}

Additional details are available on orbital-resolved (projected) band structures of LaSb primitive unit cell, 4-atoms tetragonal unit cell, free standing slab with 15-ML thick; spatial distribution of spin-resolved wavefunctions of edge states in 3-ML thick, 50-ML wide LaSb ribbon; the evolution of the topology of electronic structures in LaSb thin films having 5-ML and 7-ML thick under biaxial tensile strain; and electronic band structure of even and odd number of ML in the whole Brillouin zone. This material is available free of charge via the ACS Publications website at DOI: http://pubs.acs.org.

\end{suppinfo}

\bibliography{LaSb_QCE_DaiQHo_NanoLett}

\providecommand{\latin}[1]{#1}
\makeatletter
\providecommand{\doi}
  {\begingroup\let\do\@makeother\dospecials
  \catcode`\{=1 \catcode`\}=2 \doi@aux}
\providecommand{\doi@aux}[1]{\endgroup\texttt{#1}}
\makeatother
\providecommand*\mcitethebibliography{\thebibliography}
\csname @ifundefined\endcsname{endmcitethebibliography}
  {\let\endmcitethebibliography\endthebibliography}{}
\begin{mcitethebibliography}{62}
\providecommand*\natexlab[1]{#1}
\providecommand*\mciteSetBstSublistMode[1]{}
\providecommand*\mciteSetBstMaxWidthForm[2]{}
\providecommand*\mciteBstWouldAddEndPuncttrue
  {\def\EndOfBibitem{\unskip.}}
\providecommand*\mciteBstWouldAddEndPunctfalse
  {\let\EndOfBibitem\relax}
\providecommand*\mciteSetBstMidEndSepPunct[3]{}
\providecommand*\mciteSetBstSublistLabelBeginEnd[3]{}
\providecommand*\EndOfBibitem{}
\mciteSetBstSublistMode{f}
\mciteSetBstMaxWidthForm{subitem}{(\alph{mcitesubitemcount})}
\mciteSetBstSublistLabelBeginEnd
  {\mcitemaxwidthsubitemform\space}
  {\relax}
  {\relax}

\bibitem[Palmstr{\o}m \latin{et~al.}(1988)Palmstr{\o}m, Tabatabaie, and
  Allen~Jr]{palmstrom1988epitaxial}
Palmstr{\o}m,~C.; Tabatabaie,~N.; Allen~Jr,~S. Epitaxial growth of ErAs on
  (100) GaAs. \emph{Applied Physics Letters} \textbf{1988}, \emph{53},
  2608--2610\relax
\mciteBstWouldAddEndPuncttrue
\mciteSetBstMidEndSepPunct{\mcitedefaultmidpunct}
{\mcitedefaultendpunct}{\mcitedefaultseppunct}\relax
\EndOfBibitem
\bibitem[Buehl \latin{et~al.}(2010)Buehl, LeBeau, Stemmer, Scarpulla,
  Palmstr{\o}m, and Gossard]{buehl2010growth}
Buehl,~T.~E.; LeBeau,~J.~M.; Stemmer,~S.; Scarpulla,~M.~A.;
  Palmstr{\o}m,~C.~J.; Gossard,~A.~C. Growth of embedded ErAs nanorods on (4 1
  1) a and (4 1 1) b GaAs by molecular beam epitaxy. \emph{Journal of Crystal
  Growth} \textbf{2010}, \emph{312}, 2089--2092\relax
\mciteBstWouldAddEndPuncttrue
\mciteSetBstMidEndSepPunct{\mcitedefaultmidpunct}
{\mcitedefaultendpunct}{\mcitedefaultseppunct}\relax
\EndOfBibitem
\bibitem[Kawasaki \latin{et~al.}(2013)Kawasaki, Schultz, Lu, Gossard, and
  Palmstr{\o}m]{kawasaki2013surface}
Kawasaki,~J.~K.; Schultz,~B.~D.; Lu,~H.; Gossard,~A.~C.; Palmstr{\o}m,~C.~J.
  Surface-mediated tunable self-assembly of single crystal semimetallic
  ErSb/GaSb nanocomposite structures. \emph{Nano Letters} \textbf{2013},
  \emph{13}, 2895--2901\relax
\mciteBstWouldAddEndPuncttrue
\mciteSetBstMidEndSepPunct{\mcitedefaultmidpunct}
{\mcitedefaultendpunct}{\mcitedefaultseppunct}\relax
\EndOfBibitem
\bibitem[Hanson \latin{et~al.}(2006)Hanson, Gossard, and Brown]{hanson2006eras}
Hanson,~M.; Gossard,~A.; Brown,~E. ErAs as a transparent contact at 1.55
  $\mu$m. \emph{Applied Physics Letters} \textbf{2006}, \emph{89}, 111908\relax
\mciteBstWouldAddEndPuncttrue
\mciteSetBstMidEndSepPunct{\mcitedefaultmidpunct}
{\mcitedefaultendpunct}{\mcitedefaultseppunct}\relax
\EndOfBibitem
\bibitem[Liu \latin{et~al.}(2011)Liu, Ramu, Bowers, Palmstr{\o}m, Burke, Lu,
  and Gossard]{liu2011properties}
Liu,~X.; Ramu,~A.; Bowers,~J.~E.; Palmstr{\o}m,~C.; Burke,~P.~G.; Lu,~H.;
  Gossard,~A.~C. Properties of molecular beam epitaxially grown ScAs: InGaAs
  and ErAs: InGaAs nanocomposites for thermoelectricapplications. \emph{Journal
  of Crystal Growth} \textbf{2011}, \emph{316}, 56--59\relax
\mciteBstWouldAddEndPuncttrue
\mciteSetBstMidEndSepPunct{\mcitedefaultmidpunct}
{\mcitedefaultendpunct}{\mcitedefaultseppunct}\relax
\EndOfBibitem
\bibitem[Kawasaki \latin{et~al.}(2011)Kawasaki, Timm, Delaney, Lundgren,
  Mikkelsen, and Palmstr{\o}m]{kawasaki2011local}
Kawasaki,~J.~K.; Timm,~R.; Delaney,~K.~T.; Lundgren,~E.; Mikkelsen,~A.;
  Palmstr{\o}m,~C.~J. Local density of states and interface effects in
  semimetallic ErAs nanoparticles embedded in GaAs. \emph{Physical Review
  Letters} \textbf{2011}, \emph{107}, 036806\relax
\mciteBstWouldAddEndPuncttrue
\mciteSetBstMidEndSepPunct{\mcitedefaultmidpunct}
{\mcitedefaultendpunct}{\mcitedefaultseppunct}\relax
\EndOfBibitem
\bibitem[Kawasaki \latin{et~al.}(2013)Kawasaki, Schultz, and
  Palmstr{\o}m]{kawasaki2013size}
Kawasaki,~J.; Schultz,~B.; Palmstr{\o}m,~C. Size effects on the electronic
  structure of ErSb nanoparticles embedded in the GaSb (001) surface.
  \emph{Physical Review B} \textbf{2013}, \emph{87}, 035419\relax
\mciteBstWouldAddEndPuncttrue
\mciteSetBstMidEndSepPunct{\mcitedefaultmidpunct}
{\mcitedefaultendpunct}{\mcitedefaultseppunct}\relax
\EndOfBibitem
\bibitem[Lu \latin{et~al.}(2014)Lu, Ouellette, Preu, Watts, Zaks, Burke,
  Sherwin, and Gossard]{lu2014self}
Lu,~H.; Ouellette,~D.~G.; Preu,~S.; Watts,~J.~D.; Zaks,~B.; Burke,~P.~G.;
  Sherwin,~M.~S.; Gossard,~A.~C. Self-assembled ErSb nanostructures with
  optical applications in infrared and terahertz. \emph{Nano Letters}
  \textbf{2014}, \emph{14}, 1107--1112\relax
\mciteBstWouldAddEndPuncttrue
\mciteSetBstMidEndSepPunct{\mcitedefaultmidpunct}
{\mcitedefaultendpunct}{\mcitedefaultseppunct}\relax
\EndOfBibitem
\bibitem[Bomberger \latin{et~al.}(2017)Bomberger, Lewis, Vanderhoef, Doty, and
  Zide]{bomberger2017overview}
Bomberger,~C.~C.; Lewis,~M.~R.; Vanderhoef,~L.~R.; Doty,~M.~F.; Zide,~J.~M.
  Overview of lanthanide pnictide films and nanoparticles epitaxially
  incorporated into III-V semiconductors. \emph{Journal of Vacuum Science \&
  Technology B, Nanotechnology and Microelectronics: Materials, Processing,
  Measurement, and Phenomena} \textbf{2017}, \emph{35}, 030801\relax
\mciteBstWouldAddEndPuncttrue
\mciteSetBstMidEndSepPunct{\mcitedefaultmidpunct}
{\mcitedefaultendpunct}{\mcitedefaultseppunct}\relax
\EndOfBibitem
\bibitem[Krivoy \latin{et~al.}(2018)Krivoy, Vasudev, Rahimi, Synowicki,
  McNicholas, Ironside, Salas, Kelp, Jung, Nair, \latin{et~al.}
  others]{krivoy2018rare}
Krivoy,~E.; Vasudev,~A.; Rahimi,~S.; Synowicki,~R.; McNicholas,~K.;
  Ironside,~D.; Salas,~R.; Kelp,~G.; Jung,~D.; Nair,~H., \latin{et~al.}
  Rare-earth monopnictide alloys for tunable, epitaxial, designer plasmonics.
  \emph{ACS Photonics} \textbf{2018}, \emph{5}, 3051--3056\relax
\mciteBstWouldAddEndPuncttrue
\mciteSetBstMidEndSepPunct{\mcitedefaultmidpunct}
{\mcitedefaultendpunct}{\mcitedefaultseppunct}\relax
\EndOfBibitem
\bibitem[Tafti \latin{et~al.}(2016)Tafti, Gibson, Kushwaha, Haldolaarachchige,
  and Cava]{tafti2016resistivity}
Tafti,~F.; Gibson,~Q.; Kushwaha,~S.; Haldolaarachchige,~N.; Cava,~R.
  Resistivity plateau and extreme magnetoresistance in LaSb. \emph{Nature
  Physics} \textbf{2016}, \emph{12}, 272--277\relax
\mciteBstWouldAddEndPuncttrue
\mciteSetBstMidEndSepPunct{\mcitedefaultmidpunct}
{\mcitedefaultendpunct}{\mcitedefaultseppunct}\relax
\EndOfBibitem
\bibitem[Zeng \latin{et~al.}(2016)Zeng, Lou, Wu, Xu, Guo, Kong, Zhong, Ma, Fu,
  Richard, \latin{et~al.} others]{zeng2016LaSbcompensated}
Zeng,~L.-K.; Lou,~R.; Wu,~D.-S.; Xu,~Q.; Guo,~P.-J.; Kong,~L.-Y.; Zhong,~Y.-G.;
  Ma,~J.-Z.; Fu,~B.-B.; Richard,~P., \latin{et~al.}  Compensated semimetal LaSb
  with unsaturated magnetoresistance. \emph{Physical Review Letters}
  \textbf{2016}, \emph{117}, 127204\relax
\mciteBstWouldAddEndPuncttrue
\mciteSetBstMidEndSepPunct{\mcitedefaultmidpunct}
{\mcitedefaultendpunct}{\mcitedefaultseppunct}\relax
\EndOfBibitem
\bibitem[Wu \latin{et~al.}(2016)Wu, Kong, Wang, Johnson, Mou, Huang, Schrunk,
  Bud'ko, Canfield, and Kaminski]{wu2016asymmetric}
Wu,~Y.; Kong,~T.; Wang,~L.-L.; Johnson,~D.~D.; Mou,~D.; Huang,~L.; Schrunk,~B.;
  Bud'ko,~S.~L.; Canfield,~P.~C.; Kaminski,~A. Asymmetric mass acquisition in
  LaBi: Topological semimetal candidate. \emph{Physical Review B}
  \textbf{2016}, \emph{94}, 081108\relax
\mciteBstWouldAddEndPuncttrue
\mciteSetBstMidEndSepPunct{\mcitedefaultmidpunct}
{\mcitedefaultendpunct}{\mcitedefaultseppunct}\relax
\EndOfBibitem
\bibitem[Nayak \latin{et~al.}(2017)Nayak, Wu, Kumar, Shekhar, Singh, Fink,
  Rienks, Fecher, Parkin, Yan, \latin{et~al.} others]{nayak2017multiple}
Nayak,~J.; Wu,~S.-C.; Kumar,~N.; Shekhar,~C.; Singh,~S.; Fink,~J.;
  Rienks,~E.~E.; Fecher,~G.~H.; Parkin,~S.~S.; Yan,~B., \latin{et~al.}
  Multiple Dirac cones at the surface of the topological metal LaBi.
  \emph{Nature Communications} \textbf{2017}, \emph{8}, 1--5\relax
\mciteBstWouldAddEndPuncttrue
\mciteSetBstMidEndSepPunct{\mcitedefaultmidpunct}
{\mcitedefaultendpunct}{\mcitedefaultseppunct}\relax
\EndOfBibitem
\bibitem[Lou \latin{et~al.}(2017)Lou, Fu, Xu, Guo, Kong, Zeng, Ma, Richard,
  Fang, Huang, \latin{et~al.} others]{lou2017evidenceLaBi}
Lou,~R.; Fu,~B.-B.; Xu,~Q.; Guo,~P.-J.; Kong,~L.-Y.; Zeng,~L.-K.; Ma,~J.-Z.;
  Richard,~P.; Fang,~C.; Huang,~Y.-B., \latin{et~al.}  Evidence of topological
  insulator state in the semimetal LaBi. \emph{Physical Review B}
  \textbf{2017}, \emph{95}, 115140\relax
\mciteBstWouldAddEndPuncttrue
\mciteSetBstMidEndSepPunct{\mcitedefaultmidpunct}
{\mcitedefaultendpunct}{\mcitedefaultseppunct}\relax
\EndOfBibitem
\bibitem[Kuroda \latin{et~al.}(2018)Kuroda, Ochi, Suzuki, Hirayama, Nakayama,
  Noguchi, Bareille, Akebi, Kunisada, Muro, \latin{et~al.}
  others]{kuroda2018experimentalCePn}
Kuroda,~K.; Ochi,~M.; Suzuki,~H.; Hirayama,~M.; Nakayama,~M.; Noguchi,~R.;
  Bareille,~C.; Akebi,~S.; Kunisada,~S.; Muro,~T., \latin{et~al.}  Experimental
  determination of the topological phase diagram in cerium monopnictides.
  \emph{Physical Review Letters} \textbf{2018}, \emph{120}, 086402\relax
\mciteBstWouldAddEndPuncttrue
\mciteSetBstMidEndSepPunct{\mcitedefaultmidpunct}
{\mcitedefaultendpunct}{\mcitedefaultseppunct}\relax
\EndOfBibitem
\bibitem[Schrunk \latin{et~al.}(2022)Schrunk, Kushnirenko, Kuthanazhi, Ahn,
  Wang, O’Leary, Lee, Eaton, Fedorov, Lou, \latin{et~al.}
  others]{schrunk2022emergence}
Schrunk,~B.; Kushnirenko,~Y.; Kuthanazhi,~B.; Ahn,~J.; Wang,~L.-L.;
  O’Leary,~E.; Lee,~K.; Eaton,~A.; Fedorov,~A.; Lou,~R., \latin{et~al.}
  Emergence of Fermi arcs due to magnetic splitting in an antiferromagnet.
  \emph{Nature} \textbf{2022}, \emph{603}, 610--615\relax
\mciteBstWouldAddEndPuncttrue
\mciteSetBstMidEndSepPunct{\mcitedefaultmidpunct}
{\mcitedefaultendpunct}{\mcitedefaultseppunct}\relax
\EndOfBibitem
\bibitem[Allen~Jr \latin{et~al.}(1990)Allen~Jr, Tabatabaie, Palmstr{\o}m,
  Mounier, Hull, Sands, DeRosa, Gilchrist, and Garrison]{allen1990magneto}
Allen~Jr,~S.; Tabatabaie,~N.; Palmstr{\o}m,~C.; Mounier,~S.; Hull,~G.;
  Sands,~T.; DeRosa,~F.; Gilchrist,~H.; Garrison,~K. Magneto-transport in
  ultrathin ErAs epitaxial layers buried in GaAs. \emph{Surface Science}
  \textbf{1990}, \emph{228}, 13--15\relax
\mciteBstWouldAddEndPuncttrue
\mciteSetBstMidEndSepPunct{\mcitedefaultmidpunct}
{\mcitedefaultendpunct}{\mcitedefaultseppunct}\relax
\EndOfBibitem
\bibitem[Bernevig \latin{et~al.}(2006)Bernevig, Hughes, and
  Zhang]{bernevig2006quantum}
Bernevig,~B.~A.; Hughes,~T.~L.; Zhang,~S.-C. Quantum spin Hall effect and
  topological phase transition in HgTe quantum wells. \emph{Science}
  \textbf{2006}, \emph{314}, 1757--1761\relax
\mciteBstWouldAddEndPuncttrue
\mciteSetBstMidEndSepPunct{\mcitedefaultmidpunct}
{\mcitedefaultendpunct}{\mcitedefaultseppunct}\relax
\EndOfBibitem
\bibitem[Konig \latin{et~al.}(2007)Konig, Wiedmann, Brune, Roth, Buhmann,
  Molenkamp, Qi, and Zhang]{konig2007quantum}
Konig,~M.; Wiedmann,~S.; Brune,~C.; Roth,~A.; Buhmann,~H.; Molenkamp,~L.~W.;
  Qi,~X.-L.; Zhang,~S.-C. Quantum spin Hall insulator state in HgTe quantum
  wells. \emph{Science} \textbf{2007}, \emph{318}, 766--770\relax
\mciteBstWouldAddEndPuncttrue
\mciteSetBstMidEndSepPunct{\mcitedefaultmidpunct}
{\mcitedefaultendpunct}{\mcitedefaultseppunct}\relax
\EndOfBibitem
\bibitem[Qian \latin{et~al.}(2014)Qian, Liu, Fu, and Li]{qian2014quantum}
Qian,~X.; Liu,~J.; Fu,~L.; Li,~J. Quantum spin Hall effect in two-dimensional
  transition metal dichalcogenides. \emph{Science} \textbf{2014}, \emph{346},
  1344--1347\relax
\mciteBstWouldAddEndPuncttrue
\mciteSetBstMidEndSepPunct{\mcitedefaultmidpunct}
{\mcitedefaultendpunct}{\mcitedefaultseppunct}\relax
\EndOfBibitem
\bibitem[Tang \latin{et~al.}(2017)Tang, Zhang, Wong, Pedramrazi, Tsai, Jia,
  Moritz, Claassen, Ryu, Kahn, \latin{et~al.} others]{tang2017quantum}
Tang,~S.; Zhang,~C.; Wong,~D.; Pedramrazi,~Z.; Tsai,~H.-Z.; Jia,~C.;
  Moritz,~B.; Claassen,~M.; Ryu,~H.; Kahn,~S., \latin{et~al.}  Quantum spin
  Hall state in monolayer 1T'-WTe$_2$. \emph{Nature Physics} \textbf{2017},
  \emph{13}, 683--687\relax
\mciteBstWouldAddEndPuncttrue
\mciteSetBstMidEndSepPunct{\mcitedefaultmidpunct}
{\mcitedefaultendpunct}{\mcitedefaultseppunct}\relax
\EndOfBibitem
\bibitem[Fei \latin{et~al.}(2017)Fei, Palomaki, Wu, Zhao, Cai, Sun, Nguyen,
  Finney, Xu, and Cobden]{fei2017edge}
Fei,~Z.; Palomaki,~T.; Wu,~S.; Zhao,~W.; Cai,~X.; Sun,~B.; Nguyen,~P.;
  Finney,~J.; Xu,~X.; Cobden,~D.~H. Edge conduction in monolayer WTe$_2$.
  \emph{Nature Physics} \textbf{2017}, \emph{13}, 677--682\relax
\mciteBstWouldAddEndPuncttrue
\mciteSetBstMidEndSepPunct{\mcitedefaultmidpunct}
{\mcitedefaultendpunct}{\mcitedefaultseppunct}\relax
\EndOfBibitem
\bibitem[Wu \latin{et~al.}(2018)Wu, Fatemi, Gibson, Watanabe, Taniguchi, Cava,
  and Jarillo-Herrero]{wu2018observation}
Wu,~S.; Fatemi,~V.; Gibson,~Q.~D.; Watanabe,~K.; Taniguchi,~T.; Cava,~R.~J.;
  Jarillo-Herrero,~P. Observation of the quantum spin Hall effect up to 100
  kelvin in a monolayer crystal. \emph{Science} \textbf{2018}, \emph{359},
  76--79\relax
\mciteBstWouldAddEndPuncttrue
\mciteSetBstMidEndSepPunct{\mcitedefaultmidpunct}
{\mcitedefaultendpunct}{\mcitedefaultseppunct}\relax
\EndOfBibitem
\bibitem[Tang \latin{et~al.}(2014)Tang, Chen, Cao, Huang, Cahangirov, Xian, Xu,
  Zhang, Duan, and Rubio]{tang2014stable}
Tang,~P.; Chen,~P.; Cao,~W.; Huang,~H.; Cahangirov,~S.; Xian,~L.; Xu,~Y.;
  Zhang,~S.-C.; Duan,~W.; Rubio,~A. Stable two-dimensional dumbbell stanene: A
  quantum spin Hall insulator. \emph{Physical Review B} \textbf{2014},
  \emph{90}, 121408\relax
\mciteBstWouldAddEndPuncttrue
\mciteSetBstMidEndSepPunct{\mcitedefaultmidpunct}
{\mcitedefaultendpunct}{\mcitedefaultseppunct}\relax
\EndOfBibitem
\bibitem[Liu \latin{et~al.}(2017)Liu, Wang, Fang, Fu, and Qian]{liu2017van}
Liu,~J.; Wang,~H.; Fang,~C.; Fu,~L.; Qian,~X. van der Waals stacking-induced
  topological phase transition in layered ternary transition metal
  chalcogenides. \emph{Nano Letters} \textbf{2017}, \emph{17}, 467--475\relax
\mciteBstWouldAddEndPuncttrue
\mciteSetBstMidEndSepPunct{\mcitedefaultmidpunct}
{\mcitedefaultendpunct}{\mcitedefaultseppunct}\relax
\EndOfBibitem
\bibitem[Kandrai \latin{et~al.}(2020)Kandrai, Vancs{\'o}, Kukucska, Koltai,
  Baranka, Hoffmann, Pekker, Kamar{\'a}s, Horv{\'a}th, Vymazalov{\'a},
  \latin{et~al.} others]{kandrai2020signature}
Kandrai,~K.; Vancs{\'o},~P.; Kukucska,~G.; Koltai,~J.; Baranka,~G.;
  Hoffmann,~{\'A}.; Pekker,~{\'A}.; Kamar{\'a}s,~K.; Horv{\'a}th,~Z.~E.;
  Vymazalov{\'a},~A., \latin{et~al.}  Signature of large-gap quantum spin Hall
  state in the layered mineral jacutingaite. \emph{Nano Letters} \textbf{2020},
  \emph{20}, 5207--5213\relax
\mciteBstWouldAddEndPuncttrue
\mciteSetBstMidEndSepPunct{\mcitedefaultmidpunct}
{\mcitedefaultendpunct}{\mcitedefaultseppunct}\relax
\EndOfBibitem
\bibitem[Khalid \latin{et~al.}(2020)Khalid, Sharan, and Janotti]{Shoaib2020}
Khalid,~S.; Sharan,~A.; Janotti,~A. Hybrid functional calculations of
  electronic structure and carrier densities in rare-earth monopnictides.
  \emph{Physical Review B} \textbf{2020}, \emph{101}, 125105\relax
\mciteBstWouldAddEndPuncttrue
\mciteSetBstMidEndSepPunct{\mcitedefaultmidpunct}
{\mcitedefaultendpunct}{\mcitedefaultseppunct}\relax
\EndOfBibitem
\bibitem[Chua and Pratt(1974)Chua, and Pratt]{chua1974simple}
Chua,~K.; Pratt,~J. A simple direct-reaction calorimeter and some observations
  on the heats of formation of IIIA-VB sodium chloride structures.
  \emph{Thermochimica Acta} \textbf{1974}, \emph{8}, 409--421\relax
\mciteBstWouldAddEndPuncttrue
\mciteSetBstMidEndSepPunct{\mcitedefaultmidpunct}
{\mcitedefaultendpunct}{\mcitedefaultseppunct}\relax
\EndOfBibitem
\bibitem[Chang \latin{et~al.}(2009)Chang, Kim, Phark, Kim, Yu, and
  Noh]{chang2009fundamental}
Chang,~Y.~J.; Kim,~C.~H.; Phark,~S.-H.; Kim,~Y.; Yu,~J.; Noh,~T. Fundamental
  thickness limit of itinerant ferromagnetic SrRuO$_3$ thin films.
  \emph{Physical Review Letters} \textbf{2009}, \emph{103}, 057201\relax
\mciteBstWouldAddEndPuncttrue
\mciteSetBstMidEndSepPunct{\mcitedefaultmidpunct}
{\mcitedefaultendpunct}{\mcitedefaultseppunct}\relax
\EndOfBibitem
\bibitem[Sohn \latin{et~al.}(2021)Sohn, Kim, Kim, Lee, Hahn, Kim, Huh, Kim,
  Kim, Kyung, \latin{et~al.} others]{sohn2021observation}
Sohn,~B.; Kim,~J.~R.; Kim,~C.~H.; Lee,~S.; Hahn,~S.; Kim,~Y.; Huh,~S.; Kim,~D.;
  Kim,~Y.; Kyung,~W., \latin{et~al.}  Observation of metallic electronic
  structure in a single-atomic-layer oxide. \emph{Nature Communications}
  \textbf{2021}, \emph{12}, 1--8\relax
\mciteBstWouldAddEndPuncttrue
\mciteSetBstMidEndSepPunct{\mcitedefaultmidpunct}
{\mcitedefaultendpunct}{\mcitedefaultseppunct}\relax
\EndOfBibitem
\bibitem[Kim \latin{et~al.}(2018)Kim, Kim, and Vanderbilt]{kim2018nearly}
Kim,~J.; Kim,~H.-S.; Vanderbilt,~D. Nearly triple nodal point topological phase
  in half-metallic GdN. \emph{Physical Review B} \textbf{2018}, \emph{98},
  155122\relax
\mciteBstWouldAddEndPuncttrue
\mciteSetBstMidEndSepPunct{\mcitedefaultmidpunct}
{\mcitedefaultendpunct}{\mcitedefaultseppunct}\relax
\EndOfBibitem
\bibitem[Li \latin{et~al.}(2015)Li, Kim, Kioussis, Ning, Su, Iitaka, Tohyama,
  Yang, and Zhang]{li2015gdn}
Li,~Z.; Kim,~J.; Kioussis,~N.; Ning,~S.-Y.; Su,~H.; Iitaka,~T.; Tohyama,~T.;
  Yang,~X.; Zhang,~J.-X. GdN thin film: Chern insulating state on square
  lattice. \emph{Physical Review B} \textbf{2015}, \emph{92}, 201303\relax
\mciteBstWouldAddEndPuncttrue
\mciteSetBstMidEndSepPunct{\mcitedefaultmidpunct}
{\mcitedefaultendpunct}{\mcitedefaultseppunct}\relax
\EndOfBibitem
\bibitem[Kane and Mele(2005)Kane, and Mele]{kane2005quantum}
Kane,~C.~L.; Mele,~E.~J. Quantum spin Hall effect in graphene. \emph{Physical
  Review Letters} \textbf{2005}, \emph{95}, 226801\relax
\mciteBstWouldAddEndPuncttrue
\mciteSetBstMidEndSepPunct{\mcitedefaultmidpunct}
{\mcitedefaultendpunct}{\mcitedefaultseppunct}\relax
\EndOfBibitem
\bibitem[Sheng \latin{et~al.}(2005)Sheng, Sheng, Ting, and
  Haldane]{sheng2005nondissipative}
Sheng,~L.; Sheng,~D.; Ting,~C.; Haldane,~F. Nondissipative spin Hall effect via
  quantized edge transport. \emph{Physical Review Letters} \textbf{2005},
  \emph{95}, 136602\relax
\mciteBstWouldAddEndPuncttrue
\mciteSetBstMidEndSepPunct{\mcitedefaultmidpunct}
{\mcitedefaultendpunct}{\mcitedefaultseppunct}\relax
\EndOfBibitem
\bibitem[Fu and Kane(2007)Fu, and Kane]{fu2007topological}
Fu,~L.; Kane,~C.~L. Topological insulators with inversion symmetry.
  \emph{Physical Review B} \textbf{2007}, \emph{76}, 045302\relax
\mciteBstWouldAddEndPuncttrue
\mciteSetBstMidEndSepPunct{\mcitedefaultmidpunct}
{\mcitedefaultendpunct}{\mcitedefaultseppunct}\relax
\EndOfBibitem
\bibitem[Fukui \latin{et~al.}(2005)Fukui, Hatsugai, and Suzuki]{fukui2005chern}
Fukui,~T.; Hatsugai,~Y.; Suzuki,~H. Chern numbers in discretized Brillouin
  zone: efficient method of computing (spin) Hall conductances. \emph{Journal
  of the Physical Society of Japan} \textbf{2005}, \emph{74}, 1674--1677\relax
\mciteBstWouldAddEndPuncttrue
\mciteSetBstMidEndSepPunct{\mcitedefaultmidpunct}
{\mcitedefaultendpunct}{\mcitedefaultseppunct}\relax
\EndOfBibitem
\bibitem[Sancho \latin{et~al.}(1985)Sancho, Sancho, Sancho, and
  Rubio]{sancho1985highly}
Sancho,~M.~L.; Sancho,~J.~L.; Sancho,~J.~L.; Rubio,~J. Highly convergent
  schemes for the calculation of bulk and surface Green functions.
  \emph{Journal of Physics F: Metal Physics} \textbf{1985}, \emph{15},
  851\relax
\mciteBstWouldAddEndPuncttrue
\mciteSetBstMidEndSepPunct{\mcitedefaultmidpunct}
{\mcitedefaultendpunct}{\mcitedefaultseppunct}\relax
\EndOfBibitem
\bibitem[Qi and Zhang(2010)Qi, and Zhang]{qi2010quantum}
Qi,~X.-L.; Zhang,~S.-C. The quantum spin Hall effect and topological
  insulators. \emph{Physics Today} \textbf{2010}, \emph{63}, 33--38\relax
\mciteBstWouldAddEndPuncttrue
\mciteSetBstMidEndSepPunct{\mcitedefaultmidpunct}
{\mcitedefaultendpunct}{\mcitedefaultseppunct}\relax
\EndOfBibitem
\bibitem[Maciejko \latin{et~al.}(2011)Maciejko, Hughes, and
  Zhang]{maciejko2011quantum}
Maciejko,~J.; Hughes,~T.~L.; Zhang,~S.-C. The quantum spin Hall effect.
  \emph{Annual Review of Condensed Matter Physics} \textbf{2011}, \emph{2},
  31--53\relax
\mciteBstWouldAddEndPuncttrue
\mciteSetBstMidEndSepPunct{\mcitedefaultmidpunct}
{\mcitedefaultendpunct}{\mcitedefaultseppunct}\relax
\EndOfBibitem
\bibitem[Khalid and Janotti(2020)Khalid, and Janotti]{khalid2020trivial}
Khalid,~S.; Janotti,~A. Trivial to nontrivial topology transition in rare-earth
  pnictides with epitaxial strain. \emph{Physical Review B} \textbf{2020},
  \emph{102}, 035151\relax
\mciteBstWouldAddEndPuncttrue
\mciteSetBstMidEndSepPunct{\mcitedefaultmidpunct}
{\mcitedefaultendpunct}{\mcitedefaultseppunct}\relax
\EndOfBibitem
\bibitem[Inbar \latin{et~al.}(2022)Inbar, Ho, Chatterjee, Engel, Khalid,
  Dempsey, Pendharkar, Chang, Nishihaya, Fedorov, \latin{et~al.}
  others]{inbar2022strain}
Inbar,~H.~S.; Ho,~D.~Q.; Chatterjee,~S.; Engel,~A.~N.; Khalid,~S.;
  Dempsey,~C.~P.; Pendharkar,~M.; Chang,~Y.~H.; Nishihaya,~S.; Fedorov,~A.~V.,
  \latin{et~al.}  Strain Tuning the Band Topology of Epitaxial GdSb Quantum
  Wells. \emph{arXiv preprint arXiv:2211.15806} \textbf{2022}, \relax
\mciteBstWouldAddEndPunctfalse
\mciteSetBstMidEndSepPunct{\mcitedefaultmidpunct}
{}{\mcitedefaultseppunct}\relax
\EndOfBibitem
\bibitem[Niu \latin{et~al.}(2020)Niu, Wang, Mao, Huang, Mokrousov, and
  Dai]{niu2020antiferromagnetic}
Niu,~C.; Wang,~H.; Mao,~N.; Huang,~B.; Mokrousov,~Y.; Dai,~Y. Antiferromagnetic
  topological insulator with nonsymmorphic protection in two dimensions.
  \emph{Physical Review Letters} \textbf{2020}, \emph{124}, 066401\relax
\mciteBstWouldAddEndPuncttrue
\mciteSetBstMidEndSepPunct{\mcitedefaultmidpunct}
{\mcitedefaultendpunct}{\mcitedefaultseppunct}\relax
\EndOfBibitem
\bibitem[Liang \latin{et~al.}(2022)Liang, Chen, Zheng, Xia, Huang, Wei, Yang,
  Chen, Zhang, Xu, \latin{et~al.} others]{liang2022approaching}
Liang,~A.; Chen,~C.; Zheng,~H.; Xia,~W.; Huang,~K.; Wei,~L.; Yang,~H.;
  Chen,~Y.; Zhang,~X.; Xu,~X., \latin{et~al.}  Approaching a Minimal
  Topological Electronic Structure in Antiferromagnetic Topological Insulator
  MnBi$_2$Te$_4$ via Surface Modification. \emph{Nano Letters} \textbf{2022},
  \emph{22}, 4307--4314\relax
\mciteBstWouldAddEndPuncttrue
\mciteSetBstMidEndSepPunct{\mcitedefaultmidpunct}
{\mcitedefaultendpunct}{\mcitedefaultseppunct}\relax
\EndOfBibitem
\bibitem[Xu \latin{et~al.}(2022)Xu, Bai, Zhou, Li, Gu, Qin, Yin, Du, Zhang,
  Zhao, \latin{et~al.} others]{xu2022evolution}
Xu,~R.; Bai,~Y.; Zhou,~J.; Li,~J.; Gu,~X.; Qin,~N.; Yin,~Z.; Du,~X.; Zhang,~Q.;
  Zhao,~W., \latin{et~al.}  Evolution of the electronic structure of ultrathin
  MnBi$_2$Te$_4$ Films. \emph{Nano Letters} \textbf{2022}, \emph{22},
  6320--6327\relax
\mciteBstWouldAddEndPuncttrue
\mciteSetBstMidEndSepPunct{\mcitedefaultmidpunct}
{\mcitedefaultendpunct}{\mcitedefaultseppunct}\relax
\EndOfBibitem
\bibitem[Inoue \latin{et~al.}(2019)Inoue, Han, Hu, Suzuki, Liu, and
  Checkelsky]{inoue2019band}
Inoue,~H.; Han,~M.; Hu,~M.; Suzuki,~T.; Liu,~J.; Checkelsky,~J.~G. Band
  engineering of a magnetic thin film rare-earth monopnictide: A platform for
  high Chern number. \emph{Physical Review Materials} \textbf{2019}, \emph{3},
  101202\relax
\mciteBstWouldAddEndPuncttrue
\mciteSetBstMidEndSepPunct{\mcitedefaultmidpunct}
{\mcitedefaultendpunct}{\mcitedefaultseppunct}\relax
\EndOfBibitem
\bibitem[Inbar \latin{et~al.}(2022)Inbar, Ho, Chatterjee, Pendharkar, Engel,
  Dong, Khalid, Chang, Guo, Fedorov, \latin{et~al.} others]{inbar2022epitaxial}
Inbar,~H.~S.; Ho,~D.~Q.; Chatterjee,~S.; Pendharkar,~M.; Engel,~A.~N.;
  Dong,~J.~T.; Khalid,~S.; Chang,~Y.~H.; Guo,~T.; Fedorov,~A.~V.,
  \latin{et~al.}  Epitaxial growth, magnetoresistance, and electronic band
  structure of GdSb magnetic semimetal films. \emph{Physical Review Materials}
  \textbf{2022}, \emph{6}, L121201\relax
\mciteBstWouldAddEndPuncttrue
\mciteSetBstMidEndSepPunct{\mcitedefaultmidpunct}
{\mcitedefaultendpunct}{\mcitedefaultseppunct}\relax
\EndOfBibitem
\bibitem[Chatterjee \latin{et~al.}(2021)Chatterjee, Khalid, Inbar, Goswami,
  Guo, Chang, Young, Fedorov, Read, Janotti, \latin{et~al.}
  others]{chatterjee2021controlling}
Chatterjee,~S.; Khalid,~S.; Inbar,~H.~S.; Goswami,~A.; Guo,~T.; Chang,~Y.-H.;
  Young,~E.; Fedorov,~A.~V.; Read,~D.; Janotti,~A., \latin{et~al.}  Controlling
  magnetoresistance by tuning semimetallicity through dimensional confinement
  and heteroepitaxy. \emph{Science Advances} \textbf{2021}, \emph{7},
  eabe8971\relax
\mciteBstWouldAddEndPuncttrue
\mciteSetBstMidEndSepPunct{\mcitedefaultmidpunct}
{\mcitedefaultendpunct}{\mcitedefaultseppunct}\relax
\EndOfBibitem
\bibitem[Hohenberg and Kohn(1964)Hohenberg, and Kohn]{Hohenberg1964}
Hohenberg,~P.; Kohn,~W. Inhomogeneous Electron Gas. \emph{Physical Review}
  \textbf{1964}, \emph{136}, B864--B871\relax
\mciteBstWouldAddEndPuncttrue
\mciteSetBstMidEndSepPunct{\mcitedefaultmidpunct}
{\mcitedefaultendpunct}{\mcitedefaultseppunct}\relax
\EndOfBibitem
\bibitem[Kohn and Sham(1965)Kohn, and Sham]{KohnSham1965}
Kohn,~W.; Sham,~L.~J. Self-Consistent Equations Including Exchange and
  Correlation Effects. \emph{Physical Review} \textbf{1965}, \emph{140},
  A1133--A1138\relax
\mciteBstWouldAddEndPuncttrue
\mciteSetBstMidEndSepPunct{\mcitedefaultmidpunct}
{\mcitedefaultendpunct}{\mcitedefaultseppunct}\relax
\EndOfBibitem
\bibitem[Kresse and Furthm{\"u}ller(1996)Kresse, and
  Furthm{\"u}ller]{kresse1996efficient}
Kresse,~G.; Furthm{\"u}ller,~J. Efficient iterative schemes for ab initio
  total-energy calculations using a plane-wave basis set. \emph{Physical Review
  B} \textbf{1996}, \emph{54}, 11169\relax
\mciteBstWouldAddEndPuncttrue
\mciteSetBstMidEndSepPunct{\mcitedefaultmidpunct}
{\mcitedefaultendpunct}{\mcitedefaultseppunct}\relax
\EndOfBibitem
\bibitem[Heyd \latin{et~al.}(2003)Heyd, Scuseria, and Ernzerhof]{hse06-2003}
Heyd,~J.; Scuseria,~G.~E.; Ernzerhof,~M. Hybrid functionals based on a screened
  Coulomb potential. \emph{The Journal of Chemical Physics} \textbf{2003},
  \emph{118}, 8207--8215\relax
\mciteBstWouldAddEndPuncttrue
\mciteSetBstMidEndSepPunct{\mcitedefaultmidpunct}
{\mcitedefaultendpunct}{\mcitedefaultseppunct}\relax
\EndOfBibitem
\bibitem[Krukau \latin{et~al.}(2006)Krukau, Vydrov, Izmaylov, and
  Scuseria]{hse06}
Krukau,~A.~V.; Vydrov,~O.~A.; Izmaylov,~A.~F.; Scuseria,~G.~E. Influence of the
  exchange screening parameter on the performance of screened hybrid
  functionals. \emph{The Journal of Chemical Physics} \textbf{2006},
  \emph{125}, 224106\relax
\mciteBstWouldAddEndPuncttrue
\mciteSetBstMidEndSepPunct{\mcitedefaultmidpunct}
{\mcitedefaultendpunct}{\mcitedefaultseppunct}\relax
\EndOfBibitem
\bibitem[Perdew \latin{et~al.}(1996)Perdew, Burke, and
  Ernzerhof]{Perdew1996-PBE}
Perdew,~J.~P.; Burke,~K.; Ernzerhof,~M. Generalized Gradient Approximation Made
  Simple. \emph{Physical Review Letters} \textbf{1996}, \emph{77}, 3865\relax
\mciteBstWouldAddEndPuncttrue
\mciteSetBstMidEndSepPunct{\mcitedefaultmidpunct}
{\mcitedefaultendpunct}{\mcitedefaultseppunct}\relax
\EndOfBibitem
\bibitem[Perdew \latin{et~al.}(1997)Perdew, Burke, and
  Ernzerhof]{Perdew1997-PBE-Erratum}
Perdew,~J.~P.; Burke,~K.; Ernzerhof,~M. Generalized Gradient Approximation Made
  Simple [Phys. Rev. Lett. 77, 3865 (1996)]. \emph{Physical Review Letters}
  \textbf{1997}, \emph{78}, 1396\relax
\mciteBstWouldAddEndPuncttrue
\mciteSetBstMidEndSepPunct{\mcitedefaultmidpunct}
{\mcitedefaultendpunct}{\mcitedefaultseppunct}\relax
\EndOfBibitem
\bibitem[Ozaki(2003)]{Ozaki2003}
Ozaki,~T. Variationally optimized atomic orbitals for large-scale electronic
  structures. \emph{Physical Review B} \textbf{2003}, \emph{67}, 155108\relax
\mciteBstWouldAddEndPuncttrue
\mciteSetBstMidEndSepPunct{\mcitedefaultmidpunct}
{\mcitedefaultendpunct}{\mcitedefaultseppunct}\relax
\EndOfBibitem
\bibitem[Ozaki and Kino(2004)Ozaki, and Kino]{Ozaki2004}
Ozaki,~T.; Kino,~H. Numerical atomic basis orbitals from H to Kr.
  \emph{Physical Review B} \textbf{2004}, \emph{69}, 195113\relax
\mciteBstWouldAddEndPuncttrue
\mciteSetBstMidEndSepPunct{\mcitedefaultmidpunct}
{\mcitedefaultendpunct}{\mcitedefaultseppunct}\relax
\EndOfBibitem
\bibitem[Ozaki and Kino(2005)Ozaki, and Kino]{Ozaki2005}
Ozaki,~T.; Kino,~H. Efficient projector expansion for the ab initio LCAO
  method. \emph{Physical Review B} \textbf{2005}, \emph{72}, 045121\relax
\mciteBstWouldAddEndPuncttrue
\mciteSetBstMidEndSepPunct{\mcitedefaultmidpunct}
{\mcitedefaultendpunct}{\mcitedefaultseppunct}\relax
\EndOfBibitem
\bibitem[Gao \latin{et~al.}(2021)Gao, Wu, Persson, and Wang]{gao2021irvsp}
Gao,~J.; Wu,~Q.; Persson,~C.; Wang,~Z. Irvsp: to obtain irreducible
  representations of electronic states in the VASP. \emph{Computer Physics
  Communications} \textbf{2021}, \emph{261}, 107760\relax
\mciteBstWouldAddEndPuncttrue
\mciteSetBstMidEndSepPunct{\mcitedefaultmidpunct}
{\mcitedefaultendpunct}{\mcitedefaultseppunct}\relax
\EndOfBibitem
\bibitem[Pizzi \latin{et~al.}(2020)Pizzi, Vitale, Arita, Bl{\"u}gel, Freimuth,
  G{\'e}ranton, Gibertini, Gresch, Johnson, Koretsune, \latin{et~al.}
  others]{pizzi2020wannier90}
Pizzi,~G.; Vitale,~V.; Arita,~R.; Bl{\"u}gel,~S.; Freimuth,~F.;
  G{\'e}ranton,~G.; Gibertini,~M.; Gresch,~D.; Johnson,~C.; Koretsune,~T.,
  \latin{et~al.}  Wannier90 as a community code: new features and applications.
  \emph{Journal of Physics: Condensed Matter} \textbf{2020}, \emph{32},
  165902\relax
\mciteBstWouldAddEndPuncttrue
\mciteSetBstMidEndSepPunct{\mcitedefaultmidpunct}
{\mcitedefaultendpunct}{\mcitedefaultseppunct}\relax
\EndOfBibitem
\bibitem[Wu \latin{et~al.}(2018)Wu, Zhang, Song, Troyer, and
  Soluyanov]{wu2018wanniertools}
Wu,~Q.; Zhang,~S.; Song,~H.-F.; Troyer,~M.; Soluyanov,~A.~A. WannierTools: An
  open-source software package for novel topological materials. \emph{Computer
  Physics Communications} \textbf{2018}, \emph{224}, 405--416\relax
\mciteBstWouldAddEndPuncttrue
\mciteSetBstMidEndSepPunct{\mcitedefaultmidpunct}
{\mcitedefaultendpunct}{\mcitedefaultseppunct}\relax
\EndOfBibitem
\end{mcitethebibliography}

\end{document}





\begin{figure}
\includegraphics[width=1\textwidth]{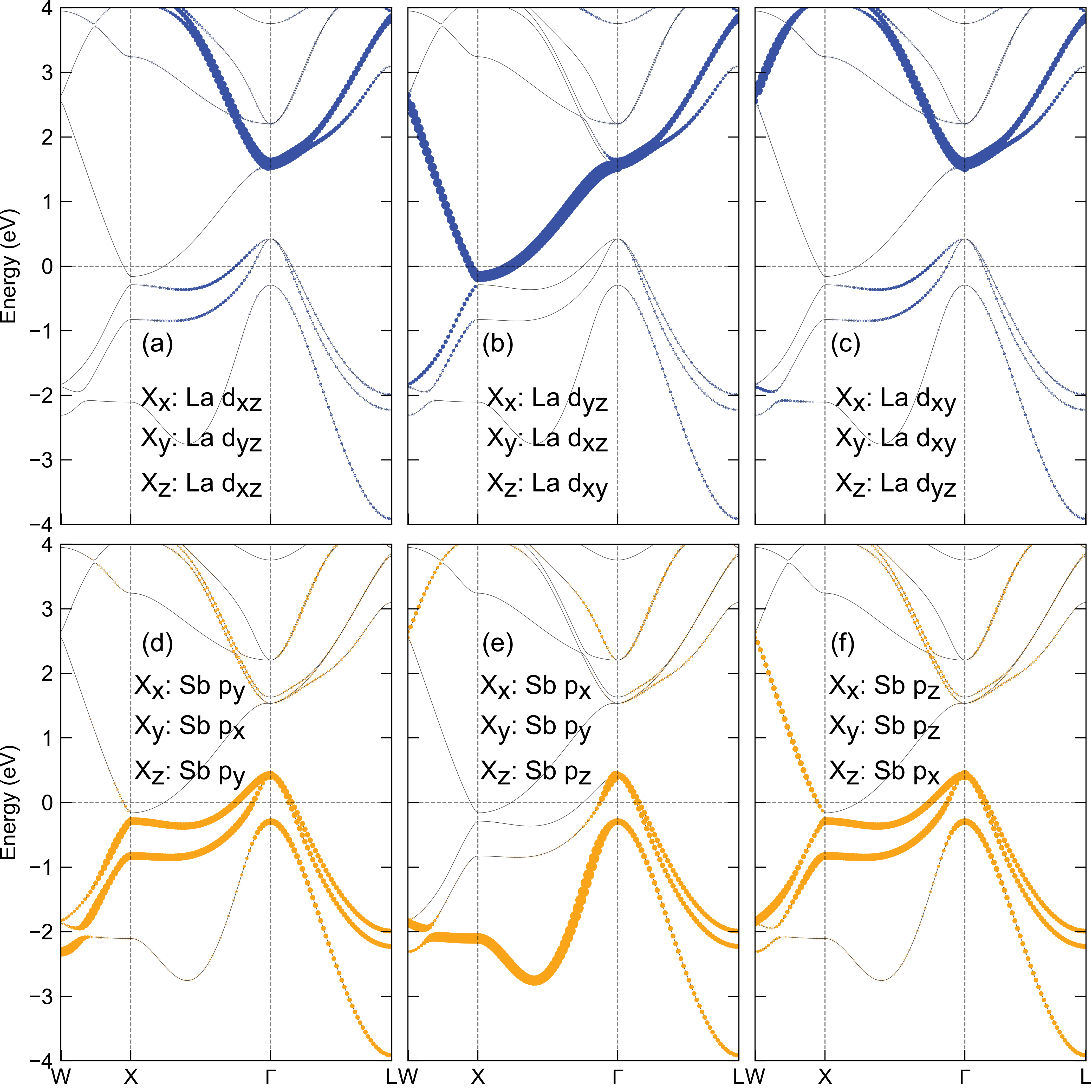}
\caption{\label{figS1} Orbital-projected band structure of LaSb bulk primitive unit cell. Due to the anisotropy of the Fermi surface, the orbital composition is different at X$_x$, X$_y$, X$_z$ as annotated in the figure above.}
\end{figure}

\begin{figure}
\centering
\includegraphics[width=1.0\textwidth]{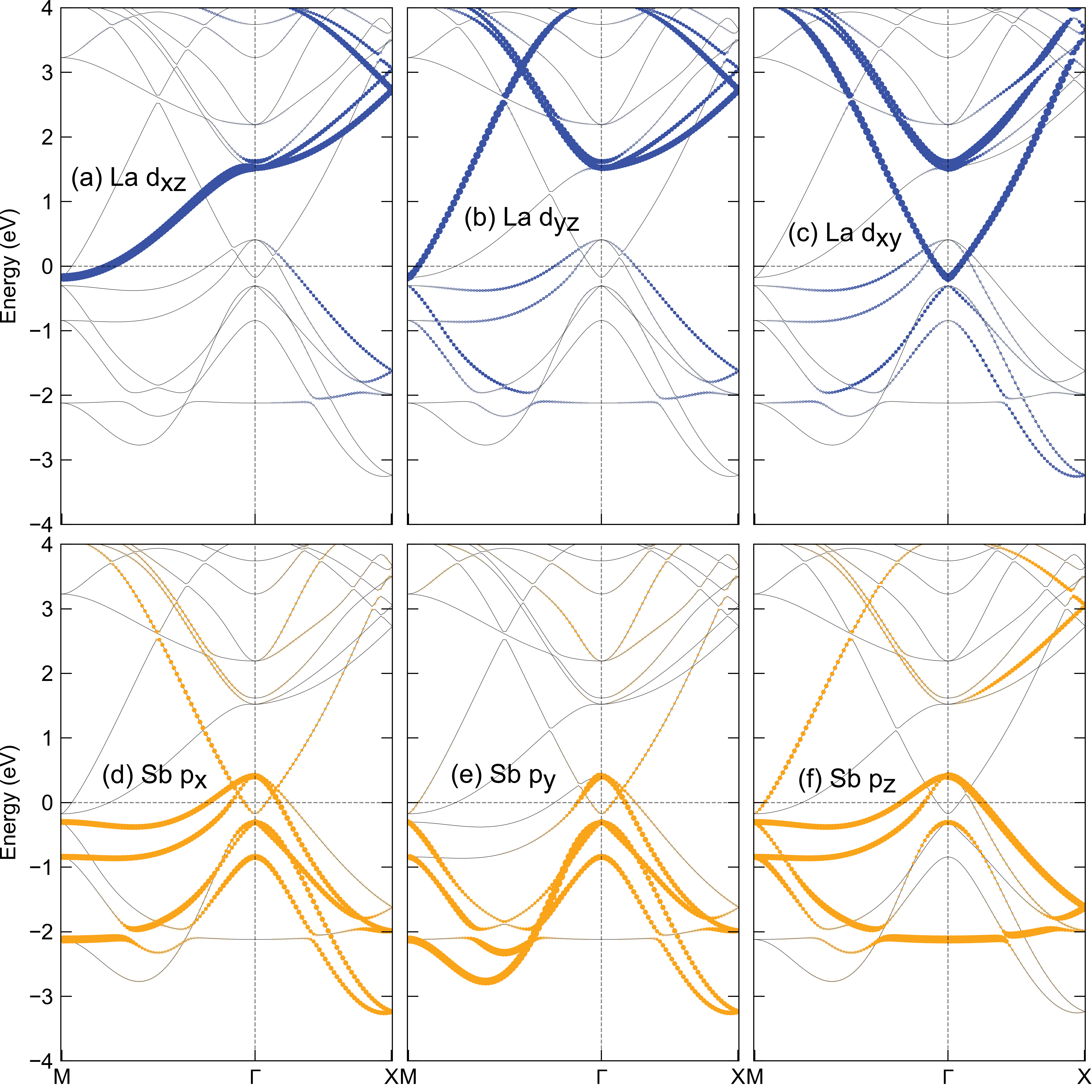}
\caption{\label{figS2} Orbital-projected band structure of LaSb bulk tetragonal unit cell along M$-\Gamma-$X as shown in Figure 1c.}
\end{figure}

\begin{figure}
\centering
\includegraphics[width=1.0\textwidth]{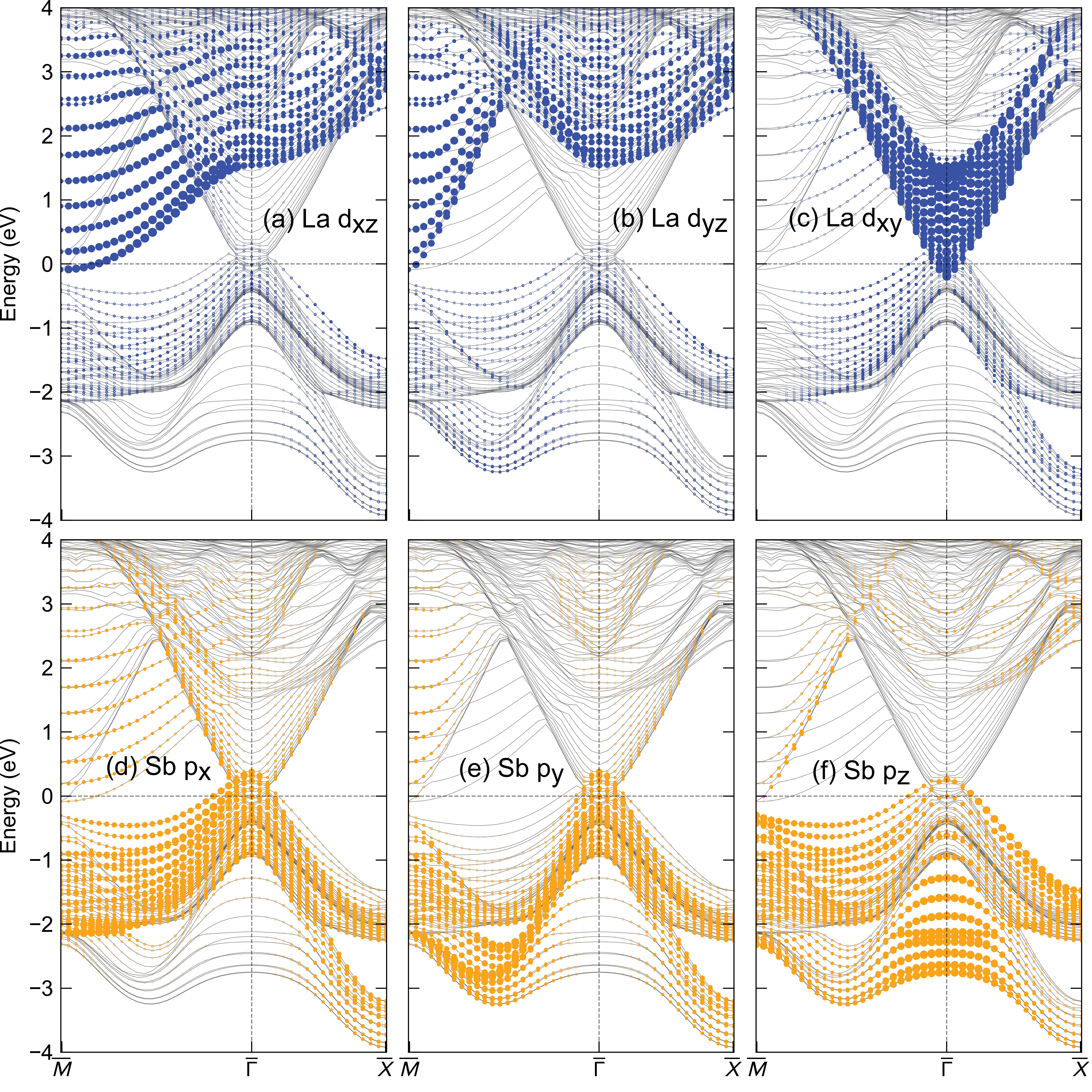}
\caption{\label{figS3} Orbital-projected band structure of 15-ML freestanding slab.}
\end{figure}

\begin{figure}
\centering
\includegraphics[width=1.0\textwidth]{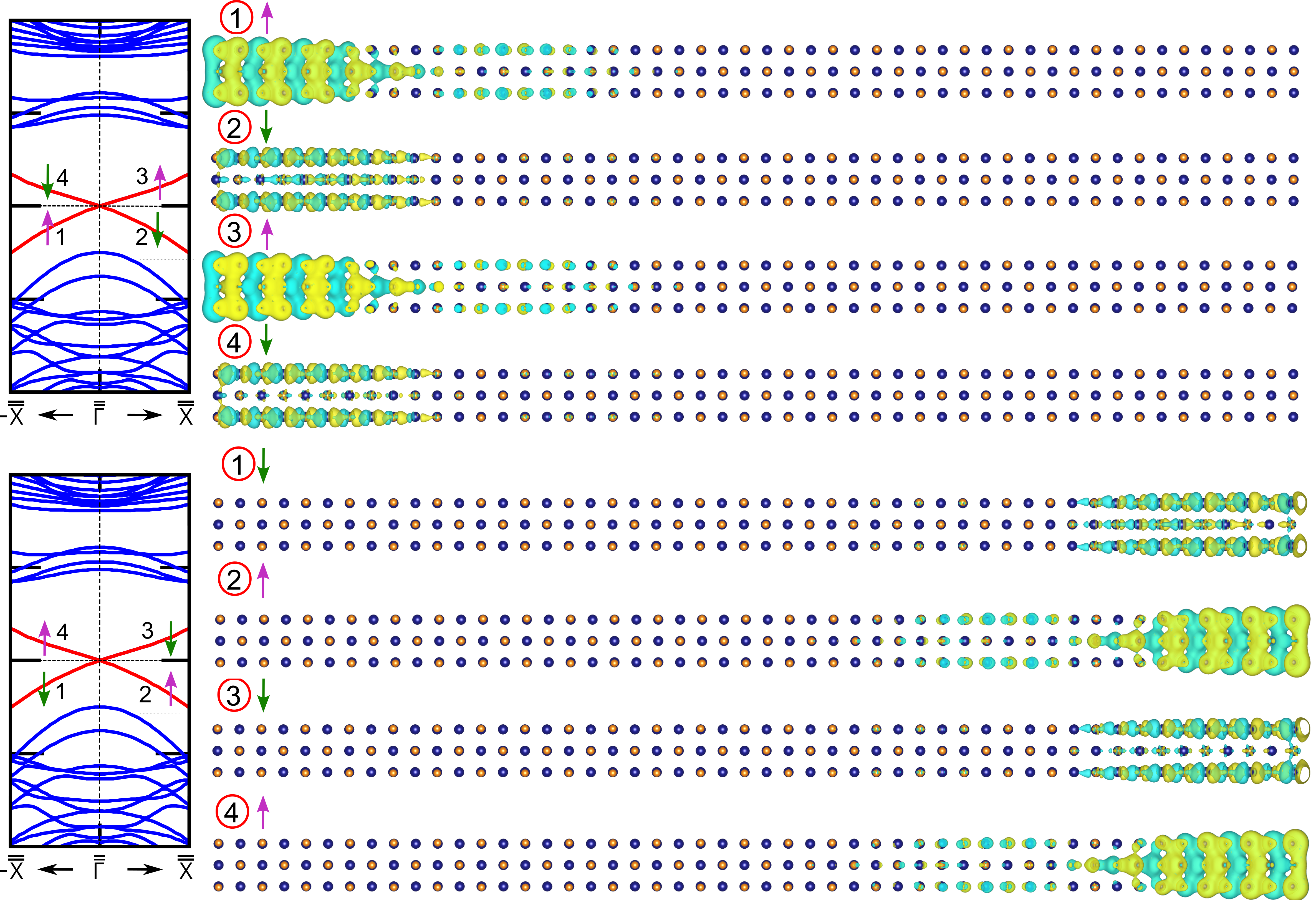}
\caption{\label{figS4} Spatial distribution of wavefunctions of edge states in LaSb 3-ML thick, 50-ML wide ribbon at the designated momentum points close to $\overline{\overline{\Gamma}}$. Views are along the periodic direction of the ribbon. Top and bottom panels show the wavefunctions associated with the left edge and the right edge, respectively. Yellow and cyan colors denote positive and negative signs of the orbitals, respectively. As can be seen here, the helical spin-momentum locking behavior is observed for both edges. On a given edge, spin up and spin down channels coexist but cannot be scattered to each other due to the presence of time-reversal symmetry.}
\end{figure}

\begin{figure}
\centering
\includegraphics[width=1.0\textwidth]{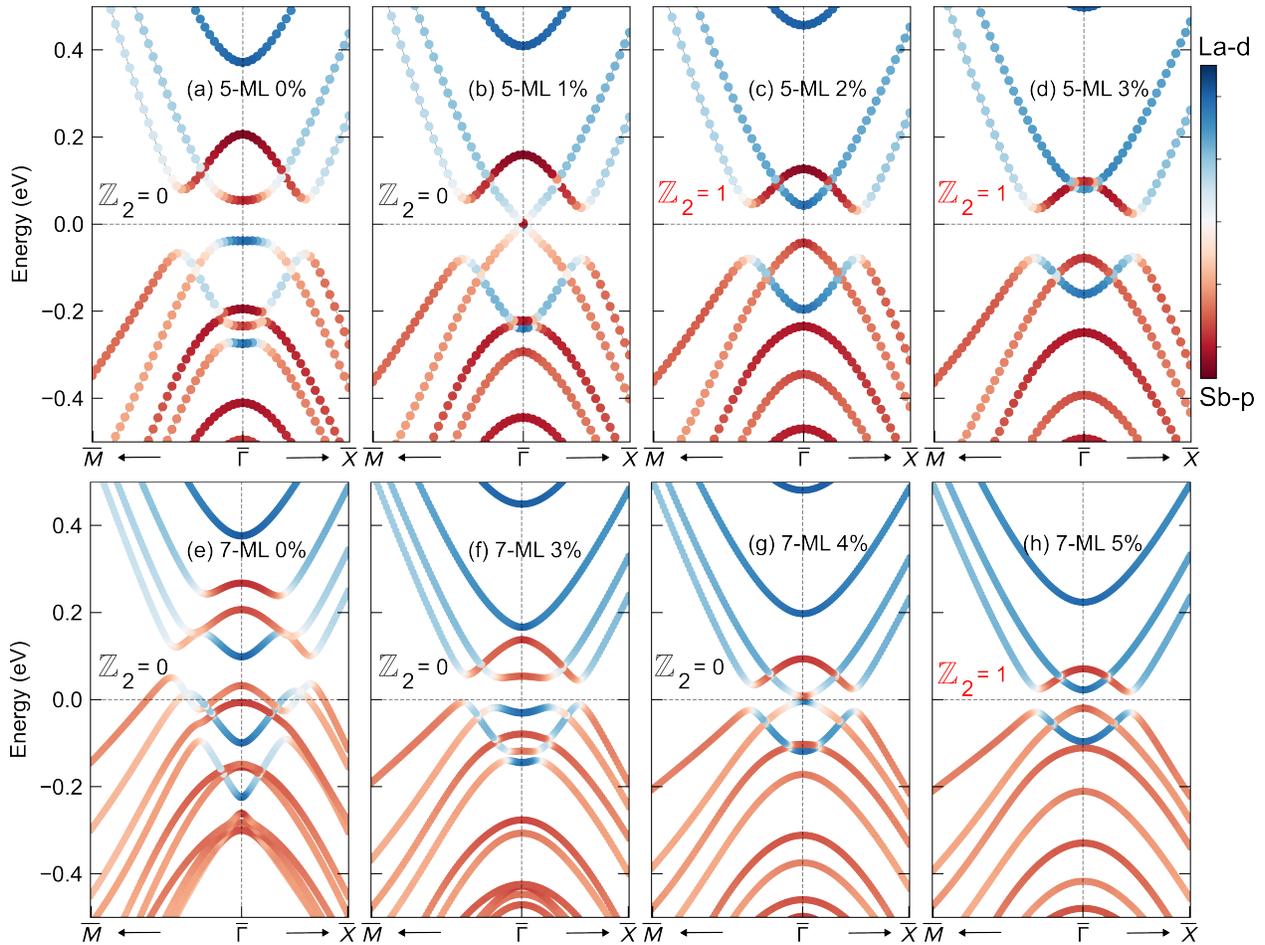}
\caption{\label{figS5}Evolution of band topology from trivial to non-trivial for 5-ML (top row) and 7-ML (bottom row) thick films under biaxial tensile strain (the applied strain levels shown in percentage). HSE06 functional was used for 5-ML slabs while SCAN functional \cite{sun2015strongly} has been used for 7-ML slabs due to computational demanding associated with large unit cells. }
\end{figure}

\begin{figure}
\centering
\includegraphics[width=1.0\textwidth]{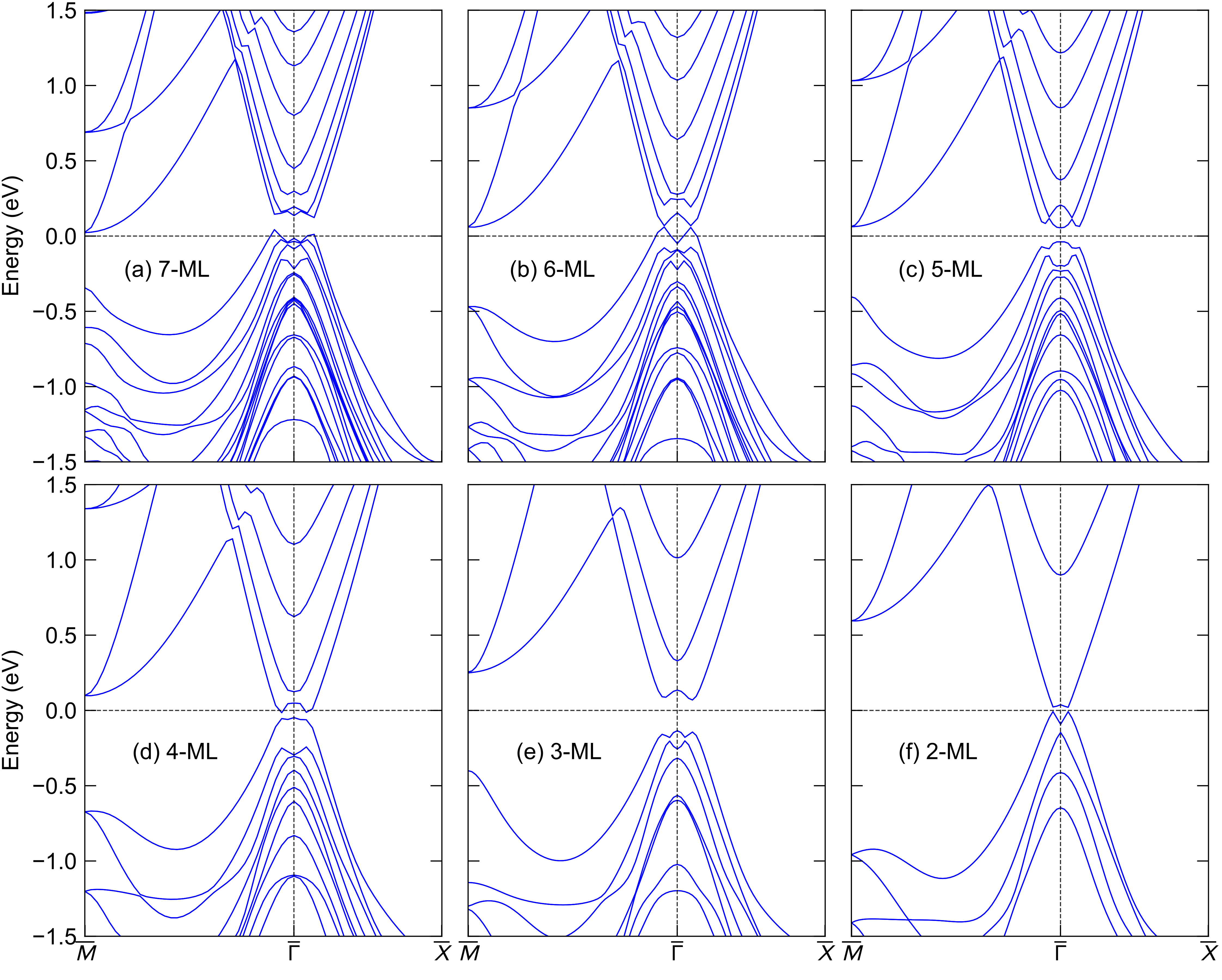}
\caption{\label{figS6} Electronic structures of freestanding  LaSb thin films with even and odd number of MLs along the high-symmetry path $\overline{\rm M}-\overline{\Gamma}-\overline{\rm X}$.}
\end{figure}


\clearpage
\bibliography{LaSb_QCE_DaiQHo_NanoLett}